\documentclass[12pt]{article}
\usepackage{amssymb,bbold}
\usepackage{epsfig}
\usepackage{bm}

\newcommand{\sect}[1]{\setcounter{equation}{0}\section{#1}}
\renewcommand{\theequation}{\arabic{section}.\arabic{equation}}

\newcommand{\e}{\epsilon}
\newcommand{\bea}{\begin{eqnarray}}
\newcommand{\eea}{\end{eqnarray}}
\newcommand{\nn}{\nonumber \\}
\newcommand{\p}[1]{(\ref{#1})}
\newcommand{\he}{{\tilde{e}}}

\newcommand{\bbeta}{{\mbox{\boldmath $\beta$}}}

\newcommand{\bchi}{{\mbox{\boldmath $\chi$}}}
\newcommand{\bomega}{{\mbox{\boldmath $\omega$}}}

\newcommand{\cross}{\times}
\newcommand{\reef}[1]{(\ref{#1})}

\def\be{\begin{equation}}
\def\ee{\end{equation}}
\def\ba{\begin{eqnarray}}
\def\ea{\end{eqnarray}}

\font\mybb=msbm10 at 11pt 
\def\bb#1{\hbox{\mybb#1}}

\def\bR {\bb{R}}

\def\bT {\bb{T}}

\def\bH {\bb{H}}

\def\cL{{\cal L}}

\def\cD{{\cal D}}

\def\cF{{\cal{F}}}

\def\e{\epsilon}

\def\de{\partial}

\begin{document}

\baselineskip 18pt

\begin{titlepage}

\vfill
\begin{flushright}
June 24 2003\\
QMUL-PH-03-08\\
hep-th/0306235\\
\end{flushright}

\vfill

\begin{center}
{\bf \Large All supersymmetric solutions of minimal supergravity }\\
\vspace*{3mm} {\bf \Large  in six dimensions}

\vskip 10.mm {Jan B. Gutowski$^{1}$, Dario Martelli$^{2}$ and Harvey S. Reall$^{3}$}
\\

\vskip 1cm

{\it
Department of Physics\\
Queen Mary, University of London\\
Mile End Rd, London E1 4NS, UK
}\\

\vspace{6pt}

\end{center}
\par

\begin{abstract}
\noindent A general form for all supersymmetric solutions of minimal
supergravity in six dimensions is obtained. Examples of new
supersymmetric solutions are presented.
It is proven that the only maximally supersymmetric
solutions are flat space, $AdS_3 \times S^3$ and a plane wave. 
As an application of the general solution, it is shown
that any supersymmetric solution with a compact
horizon must have near-horizon geometry $\bR^{1,1}
\times \bT^4$, $\bR^{1,1} \times K3$ or identified $AdS_3 \times S^3$.
\end{abstract}

\vskip 1cm
\vfill \vskip 5mm \hrule width 5.cm \vskip 5mm {\small
\noindent $^1$ E-mail: j.b.gutowski@qmul.ac.uk \\
\noindent $^2$ E-mail: d.martelli@qmul.ac.uk \\
\noindent $^3$ E-mail: h.s.reall@qmul.ac.uk}
\end{titlepage}

\sect{Introduction}

The usual approach to finding supersymmetric solutions of supergravity
theories is to make some physically motivated ansatz for the bosonic
fields, and then seek examples of this ansatz that admit one or more
supercovariantly constant ``Killing'' spinors. While this approach is often
fruitful, it would be useful to obtain a more systematic method for
finding supersymmetric solutions. In particular, given a supergravity
theory, it is natural to ask whether one can obtain {\it all}
supersymmetric solutions of that theory.

It turns out that this {\it is} possible for certain theories. Twenty
years ago, following the derivation of a BPS inequality in \cite{gibbons:82},
Tod managed to determine all supersymmetric solutions of minimal $N=2$, $D=4$
supergravity \cite{tod:83}. Starting from a Killing spinor $\epsilon$,
his strategy was to construct bosonic objects quadratic in $\epsilon$,
such as the vector $V^{\alpha} = \bar{\epsilon} \gamma^{\alpha}
\epsilon$. Fierz identities imply algebraic relations between such
quantities, and the supercovariant-constancy of $\epsilon$ yields
differential relations. For example, in this case $V$ turns out to be a
Killing vector field. Tod showed that these relations
are sufficient to fully determine the local form of the solution. The
solutions fall into two classes. In the first class, $V$ is timelike
and the solutions are the Israel-Wilson-Per\'jes solutions of
Einstein-Maxwell theory, which are specified by harmonic functions on
$\bR^3$. In the second class, $V$ is null and the
solutions are certain pp-waves, specified by harmonic functions on
$\bR^2$. Some generalizations of this result to
other $D=4$ theories were presented in \cite{tod:95}.

This method has recently been extended to certain $D=5$ theories. In
\cite{d5}, all supersymmetric solutions of minimal $D=5$ supergravity
were obtained. Once again, one can construct a Killing vector field $V$
from a Killing spinor, and the solutions fall into a ``timelike'' and
a ``null'' class. The solutions in the null class are
plane-fronted waves, and are specified in terms of harmonic functions on $\bR^3$.
The solutions in the timelike class have metric
\be
\label{eqn:d5metric}
 ds_5^2 = f^2 \left( dt + \omega \right)^2 - f^{-1} ds_4^2,
\ee
where $V = \partial/\partial t$ and
$ds_4^2$ is the line element of an arbitrary hyper-K\"ahler
4-manifold ${\cal B}$ referred to as the ``base space''. $f$ and $\omega$ are a
scalar and $1$-form on ${\cal B}$ that must obey
\be
\label{eqn:d5eqs}
 \nabla^2 f^{-1} = \frac{4}{9} \left( G^+ \right)^2, \qquad d G^+ = 0,
\ee
where $G^+$ is the self-dual part of $f d\omega$ with respect to the
metric on ${\cal B}$, and $\nabla^2$ the Laplacian on ${\cal B}$. The
solution for the gauge field is given in \cite{d5}.

The analysis of \cite{d5} shows that these purely bosonic equations
are necessary {\it and} sufficient conditions for supersymmetry. In
contrast to the null class, and the $D=4$ solutions, the general
solution to these equations is not known, so the $D=5$ timelike
solutions are determined somewhat implicitly. Ultimately, one still has to
make an ansatz for $f$ and $\omega$ to solve these equations once
${\cal B}$ has been chosen. However, this represents a
significant advance over the usual approach of making an
ansatz for the entire metric and gauge field.

Another $D=5$ theory to which this method has been applied is minimal
gauged supergravity \cite{GG}. The underlying algebraic structure
is the same as for the ungauged theory, thus one still has
the timelike and null classes to consider. However, the differential
conditions obeyed by the spinorial bi-linears are different. $V$ is still a Killing
vector so that the metric of the timelike class can be written as above,
whereas one finds that ${\cal B}$ is now a  K\"ahler manifold.
$f$ and $\omega$ are
determined essentially uniquely, although rather implicitly, once
${\cal B}$ has been chosen. The null class is specified by solving
certain nonlinear scalar equations on $\bR^3$. Although the solutions
are presented implicitly in terms of solutions of these
equations, solving these equations only involves making
ans\"atze for certain scalars rather than for the full metric and gauge
field.

This strategy of determining the solution of a supergravity theory using the
differential equations obeyed by certain spinorial bi-linears is closely related
to the mathematical notions of $G$-structures and ``intrinsic torsion''
\cite{GMPW}. Given the existence of a spinor over a $d$-dimensional manifold
$M$, the spinor will be invariant under some isotropy group $G \subset Spin(d)$.
Thus the differential forms one constructs as bilinears
enjoy the same invariance, and this defines, at least locally,
a canonical reduction of the (spin cover of the) tangent bundle $Spin (d) \to G$.
The conditions obeyed by the differential forms can be shown to be
equivalent to the Killing spinor equations.

In \cite{GP}, the techniques used to analyse the four and five
dimensional supergravities mentioned above were applied to $D=11$ supergravity. It
was shown that the existence of at least one Killing  spinor
implies that there is a Killing vector which is either timelike or null
\cite{FOF}, corresponding to a  $SU(5)$ and $(Spin(7)\ltimes
\bR^8)\ltimes \bR$ structure respectively.  The timelike case was examined in detail,
and purely bosonic necessary and sufficient conditions for preservation
of at least one supersymmetry were
obtained. As might be expected, these conditions encode the full
solution in a somewhat implicit manner. Using the same techniques,
all {\it static} solutions of the $D=10$ type II theories with only
NS flux  were analysed in \cite{GMW}. In this case,
a more refined analysis of solutions preserving different amount of
supersymmetries was presented. Static solutions always preserve more
than one supersymmetry. In the general case, when there is only one 
supersymmetry in $D=10$, there is a null Killing vector field whose
isotropy group is $Spin(7)\ltimes \bR^8$.

Given the complexity of the results in $D=10$ and $D=11$,
it appears that, as far as finding examples of new solutions is concerned,
the ``$G$-structures approach'' is only
significantly preferable to the usual ansatz-based approach in
sufficiently simple theories. There are two natural candidates for
theories in which this method might be particularly useful. The first
is minimal $N=2$, $D=4$ gauged supergravity. Understanding this case
might lead to a better understanding of the results of \cite{GG}. The
second is minimal $D=6$ supergravity since, after the minimal $N=2$, $D=4$
and $N=1$, $D=5$ theories, this is the simplest ungauged supergravity theory
with $8$ supercharges. Furthermore, it is a natural generalization of
these theories because they can be obtained from it by dimensional
reduction and truncation.

In this paper, we shall apply the methods described above to minimal $D=6$
supergravity. The bosonic sector of this theory consists of a graviton
and self-dual 3-form. One novel feature that arises in this case is that
 the Killing vector $V$ obtained from the Killing
spinor is always null and one has correspondingly an
$SU(2)\ltimes \bR^4$ structure \cite{bryant},
so there is no ``timelike'' case to consider.
The $D=6$ minimal theory arises as a consistent truncation of
higher dimensional supergravities, which is reflected in the fact that
$SU(2)\ltimes \bR^4\subset Spin(7)\ltimes \bR^8 \subset (Spin(7)\ltimes
\bR^8)\ltimes \bR$.  The solutions
can be trivially uplifted to solutions of $D=10$ and $D=11$
supergravities on flat tori.

In contrast to the $D=5$ case \cite{d5}, the null Killing vector in $D=6$ is not
hypersurface-orthogonal; this
provides the main source of complication in our equations (similar
complications would arise in general in $D=10$ and in the null case in $D=11$).
Some insight into the nature of the $D=6$ solutions can be obtained by
noting that they must contain as subsets (the oxidation of)
the timelike and null classes of the minimal $D=5$ theory.
The timelike class of the
latter theory involves an arbitrary hyper-K\"ahler manifold, and the
null class contains solutions with arbitrary dependence on a
retarded time coordinate $u$. This suggests that there should be $D=6$
solutions that exhibit both of these features, i.e., hyper-K\"ahler
spaces whose moduli are arbitrary functions of some coordinate $u$. In fact
the general supersymmetric solution with {\it vanishing flux} has precisely
this form \cite{bryant}.

It turns out that supersymmetric solutions with flux are much more
complicated. Coordinates can be introduced so that the solutions are
expressed in terms of a four dimensional $u$-dependent base manifold
${\cal B}$. In general ${\cal B}$ exhibits a {\it non-integrable} 
hyper-K\"ahler structure. Necessary and sufficient conditions for supersymmetry can
be expressed as equations for various bosonic quantities defined on
${\cal B}$. These equations are more complicated than those
encountered in $D=5$. Nevertheless, as emphasized above, it is much
easier to find supersymmetric solutions by substituting an ansatz into these
equations than it is to start with an ansatz for the entire metric and
3-form.

There are two special cases in which the solutions simplify to
yield an integrable hyper-K\"ahler structure. The first arises when
the null Killing vector field is hyper-surface orthogonal
(this include the case of vanishing flux). In this case, our solutions
are closely related to the chiral null models of
\cite{horowitz:95}. The second arises when there is no $u$-dependence,
i.e., the solution admits a second Killing vector field. In this case
the solutions are related to the generalized chiral null models of
\cite{tseytlin:96}. In this case, the necessary and sufficient
conditions for supersymmetry take a simple form similar to those for
the timelike class of the minimal $D=5$ theory and, with a few
additional assumptions, can be solved explicitly.

In the minimal $N=2$, $D=4$ theory and the minimal $D=5$ theory,
supersymmetric solutions must preserve either $1/2$ or all of the
supersymmetry, and the same is true for the minimal $D=6$ theory
(although not the minimal $D=5$ gauged theory \cite{GG}). Determining
which solutions of the $D=5$ theory are {\it maximally}
supersymmetric is rather involved \cite{d5}. Happily, this is much
easier for the $D=6$ theory, and we shall show that the only such
solutions are flat space, $AdS_3 \times S^3$ and the plane wave
solution of \cite{meesnote}.

An example of the utility of the present approach was given in
\cite{reall:02}, where the analysis of \cite{d5} was exploited to
prove a uniqueness theorem for supersymmetric black hole solutions of
minimal $D=5$ supergravity (non-supersymmetric $D=5$ black holes are not
unique \cite{emparan:02,elvang:03} unless static
\cite{gibbons:02a,gibbons:02b}). It would be interesting to see if a similar
uniqueness theorem could be proved for supersymmetric black strings in
minimal $D=6$ supergravity (non-supersymmetric black strings are not
unique \cite{horowitz:01,gubser:01,horowitz:02,wiseman:03}).
Here we shall content ourselves with
taking the first step towards such a proof, namely classifying the
possible near-horizon geometries of supersymmetric solutions with
compact horizons (e.g. a wrapped string). It turns out that there are just 3 possibilities:
$\bR^{1,1} \times \bT^4$, $\bR^{1,1} \times K3$ and $AdS_3 \times S^3$. In
the latter case, the solution must be identified so as to render the
horizon compact.

This paper is organized as follows. In section \ref{sec:basics} we
construct bosonic objects quadratic in the Killing spinor and derive
the algebraic and differential conditions they satisfy. In section
\ref{sec:gensol} we show how these conditions lead to necessary and
sufficient conditions for supersymmetry, formulated in terms of
equations on ${\cal B}$. Section \ref{Gstructures} discusses how our
work fits into the general approach of classifying supersymmetric
solutions using $G$-structures. In section \ref{sec:examples} we study
some special cases of the general solution, explain how these are
related to previous work, and construct some examples of new
solutions. Section \ref{sec:horizon} contains our classification of
possible near-horizon geometries and section \ref{sec:max} the
classication of maximally supersymmetric solutions. Finally,
section \ref{sec:outlook} contains suggestions for future work.

\sect{Minimal six-dimensional supergravity}

\label{sec:basics}

The field content of minimal six-dimensional supergravity is
the graviton $g_{\mu\nu}$, a two-form $B_{\mu\nu}^+$ with self-dual field
strength, and a symplectic Majorana-Weyl (left-handed i.e.
$\gamma_7\psi_\mu^A=-\psi_\mu^A$) gravitino $\psi_\mu^A$. $A$ is an $Sp(1)$
index which we will often suppress. This theory and the extensions
coupled to various matter multiplets are described in \cite{NS1,NS2}.

Writing a Lagrangian for this theory is notoriously complicated by
the self-duality constraints, however the addition of a tensor
matter multiplet allows one to write an action from which the
equations of motion follow. This multiplet comprises of a  two-form $B^-_{\mu\nu}$
with anti-self dual field strength, a right-handed symplectic
Majorana-Weyl field $\chi$, and a scalar field $\varphi$. The
Lagrangian, equations of motion, and supersymmetry variations can
be written in terms of the field $G_{\mu\nu\rho}=
3\de_{[\mu}B_{\nu\rho]}^+ +3\de_{[\mu}B_{\nu\rho]}^-$ and are
given for instance in \cite{NS2}.

In this paper we will be interested in the minimal theory, so we consistently
set the tensor multiplet to zero. The supersymmetry equation of the
gravitino then can be written
\bea
\label{susyeq}
\nabla_\mu \e
-\frac{1}{4}G_{\mu\rho\lambda}
\gamma^{\rho\lambda}\e & = &0
\eea
where here, and henceforth, we set $G=dB^+$. In this form, equation
(\ref{susyeq}) has a clear geometrical interpretation; namely it implies
that  $\epsilon $ is a spinor parallel with respect to a modified
spin connection $\bar\nabla$ with torsion $G$ (see also \cite{Lu:2003yt}).

The field equations are
\bea
\label{eqn:coclos}
\nabla_\mu G^{\mu\nu\rho} & = &0\\
\label{eqn:einfld}
R_{\mu\nu}  & = & G_{\mu\rho\sigma}G_\nu{}^{\rho \sigma} \label{Einst_eq+}~ .
\eea
Note that ({\ref{eqn:coclos}}) is equivalent to the Bianchi identity $dG=0$ as a
consequence of the self-duality of $G$.

Solutions of minimal $D=6$ supergravity can be trivially oxidized to
yield solutions of type II supergravity in which only NS-NS sector
fields are excited. The extra $4$ spatial dimensions $z^i$ just form a flat
torus:
\be
 ds_{10}^2 = ds_6^2 - d{\bf z}^2 \ .
\ee
In $D=10$, the NS-NS field strength $H$ is given
by $H = 2G$, so it is self-dual in the first six dimensions with
vanishing components in the torus directions.
Furthermore, the dilaton is constant. Roughly speaking, such
solutions carry equal F-string and NS5-brane charge.

\subsection*{Bi-linears and their Constraints}

We can now construct spinor bi-linears, and compute the differential
conditions they obey, in order to reexpress the supersymmetry equation in terms
of bosonic form fields defined on space-time. Mathematically, these
encode information about the underlying $G$-structure, on which we comment in
section \ref{Gstructures}. Given that
we can set the conjugation matrix $C$ to unity, we always
have $\bar{\e}^A=\e^{AT}$. The non-zero bi-linears that we have are
\bea
V_\mu \e^{AB} & = & \bar{\e}^A\gamma_\mu\e^B \label{defV}\\
\Omega^{AB}_{\mu\nu\rho} & = & \bar{\e}^A\gamma_{\mu\nu\rho}\e^B \label{defO}
\eea
while the even forms vanish pulling through a $\gamma_7$. The
three-forms are self-dual. Using \reef{reality} we can also check that the
following reality properties hold
\be
\Omega^{11*} = \Omega^{22}, \qquad \Omega^{12*} = - \Omega^{21} =  -\Omega^{12}~,
\ee
so that it will be convenient to work with three real self-dual
three-forms $X^1$, $X^2$ and $X^3$ defined by
\be
\Omega^{11} = X^1+iX^2, \qquad 
\Omega^{22} = X^1-iX^2, \qquad  
\Omega^{12} = -i X^3~.
\ee

\subsubsection*{Algebraic Constraints}

Simple algebraic relations between these bi-linears can be constructed
using the Fierz identity, as explained in Appendix \ref{conventions}. One obtains
\be
 V_\mu V^\mu=0
\label{Visnull}
\ee
so $V$ is null. We also find
\be
\label{eqn:VdotX}
i_V X^i =0
\ee
where, for a vector $Y$ and $p$-form $A$, $i_Y A$ denotes the
$(p-1)$-form obtained by contracting $Y$ with the first index of $A$.
From the self-duality of $X^i$ this is equivalent to
\be
V \wedge X^i =0~.
\ee
The 3-forms $\Omega^{AB}$ are found to obey an algebra
(\ref{eqn:compalg}). When expressed in terms of the real 3-forms $X^i$
this reads
\bea
\label{eqn:simpleralg}
X^{i \ \rho \sigma}{}_\nu X^{j \ \nu \lambda \mu} &=&\epsilon^{ijk}
\big( X^{k \ \sigma \rho \mu}V^\lambda  -X^{k \ \sigma \rho \lambda}V^\mu
\big)
\nn
&-& \delta^{ij} \big(g^{\sigma \lambda}V^\mu V^\rho + g^{\mu \rho}V^\lambda V^\sigma
-g^{\sigma \mu}V^\rho V^\lambda
-g^{\rho \lambda} V^\mu V^\sigma \big)~,
\eea
where $\epsilon^{123}=+1$.
At this point it is useful to introduce a null orthonormal basis in which
\be
\label{eqn:metsplit}
ds^2 = 2 e^+ e^- - \delta_{ab} e^a  e^b
\ee
where $e^+=V$ and $a,b=1,2,3,4$. We shall take orientation given by
\be
 \epsilon^{+-1234} = 1.
\ee
Equation \p{eqn:VdotX} and the self-duality of $X^i$ imply that
\be
\label{eqn:xsimp}
X^i = V \wedge I^i
\ee
where
\be
 I^i = {1 \over 2} I^i{}_{ab} e^a \wedge e^b.
\ee
The self-duality of $X^i$ implies that $I^i$ is anti-self dual
with respect to the metric $\delta_{ab} e^a  e^b$ with orientation
$\epsilon^{abcd} = \epsilon^{+-abcd}$.
Substituting this expression for $X^i$ into the algebra defined by
({\ref{eqn:simpleralg}}) we find that
\be
\label{eqn:quat}
(I^i)^a{}_c (I^j)^c{}_b = \epsilon^{ijk} (I^k)^a{}_b - \delta^{ij} \delta^a{}_b
\ee
where in the above, we have raised the indices of $I$ using $\delta^{ab}$.
Hence, the $I^i$ satisfy the algebra of the imaginary unit quaternions
on a 4-manifold equipped with metric $\delta_{ab} e^a e^b$.

Finally, the Fierz identity implies that the Killing spinor must obey
the projection
\be
 V \cdot \gamma \epsilon=0,
\ee
which, written in the above basis is
\be
\label{eqn:proj}
 \gamma^+ \epsilon = 0.
\ee

\subsubsection*{Differential Constraints}

The vector $V$ and 3-forms $X^i$ satisfy differential constraints
which  hold because the spinor must
satisfy the Killing spinor equation, namely it is parallel with respect to
the connection $\bar\nabla$ with torsion $G$. Explicitely, the constraint on
$V$ is
\be
\label{eqn:vdc}
\nabla_\alpha V_\beta = V^\lambda G{}_{\lambda \alpha \beta}
\ee
and the constraints on $X^i$ are
\be
\label{eqn:xdc}
\nabla_\alpha X^i_{\beta\gamma\delta}  =  G_{\alpha\beta}{}^\rho
X^i_{\rho\gamma\delta}  + G_{\alpha\gamma}{}^\rho
X^i_{\rho\delta\beta} + G_{\alpha\delta}{}^\rho
X^i_{\rho\beta\gamma}~.
\ee
In particular, we note that because $X^i$ and $G$ are self-dual,
it follows from (\ref{eqn:xdc}) that
\bea
dX^i = d^\dagger X^i = 0~. \label{Omegaclosed}
\eea
Furthermore, (\ref{eqn:vdc}) implies that $\nabla_{(\alpha} V_{\beta)}=0$, so
$V$ is a Killing vector, and also $dV = 2i_V G$. An argument in \cite{reall:02} proves that
$V$ (and $\epsilon$) cannot vanish anywhere, assuming analyticity; hence any
supersymmetric solution will admit a globally defined null Killing vector.
It is also useful to note that
\be
\label{eqn:liedx}
\cL_V X^i = \cL_V G = 0,
\ee
using $dG = 0$. Hence $V$ generates a symmetry of the full solution.

To proceed, we shall rewrite the differential constraints ({\ref{eqn:vdc}})
and ({\ref{eqn:xdc}}) in an equivalent form. Let $\omega$ denote the
spin connection. Then ({\ref{eqn:vdc}}) is equivalent to
\be
\label{eqn:spafix}
\omega_{\alpha \beta -} =G_{\alpha \beta -}.
\ee
It is straightforward to show that ({\ref{eqn:spafix}}) together with ({\ref{eqn:liedx}})
can be used to simplify ({\ref{eqn:xdc}}) and rewrite it
as an equation for $I^i$:
\be
\label{eqn:evensmp}
\nabla_\alpha I^i_{bc} = G_{\alpha b}{}^d I^i_{dc}-G_{\alpha c}{}^d I^i_{db}\ .
\ee
To summarize, the differential constraints (\ref{eqn:vdc}) and (\ref{eqn:xdc})
are equivalent to ({\ref{eqn:spafix}}) and ({\ref{eqn:evensmp}}).

\subsubsection*{The Killing spinor}

The algebraic and differential constraints on $V$ and $I^i$ are clearly
necessary conditions for the background to admit a supercovariantly
constant spinor $\epsilon$ satisfying the projection \p{eqn:proj}.
It turns out that these conditions are also sufficient.
To see this it is convenient to choose the basis vectors
$e^a$ so that the 2-forms $I^i$ have constant components. For
example, one could choose a basis so that
\be
 I^1 = e^1 \wedge e^2 - e^3 \wedge e^4, \qquad I^2 = e^1 \wedge e^3 + e^2
 \wedge e^4, \qquad I^3 = e^1\wedge e^4 - e^2\wedge e^3.
\ee
Equation ({\ref{eqn:evensmp}}) then reduces to
\be
\label{eqn:antisd}
(\omega_{\alpha a b} - G_{\alpha a b})^-=0
\ee
where here ${}^-$ denotes the anti-self-dual projection in the indices
$a$, $b$. It can then be checked that equations \p{eqn:proj},
({\ref{eqn:antisd}}) and ({\ref{eqn:spafix}}) imply that the Killing
spinor equation reduces to
\be
\label{eqn:constantspinor}
 \partial_\mu \epsilon = 0.
\ee
Hence, provided the above algebraic and differential conditions are
satisfied then, in this basis, the Killing spinor equation is
satisfied by any constant spinor obeying the projection
\p{eqn:proj}. This is the only projection so the solution must
preserve either $1/2$ or all of the supersymmetry.

In summary, the above algebraic and differential conditions on $V$ and
$I^i$ are necessary {\it and} sufficient to guarantee the
existence of a  $\bar\nabla$-parallel chiral spinor obeying
\p{eqn:proj}. Furthermore, all solutions must preserve
either $1/2$ or all of the supersymmetry.
In the next section we shall introduce coordinates and examine further
the conditions on $V$ and $I^i$ in order to obtain convenient
forms for the necessary and sufficient conditions for supersymmetry.

\sect{All supersymmetric solutions}

\label{sec:gensol}

\subsubsection*{Introduction of coordinates}

\label{subsec:coords}

Coordinates can be introduced locally as follows.
Pick a hypersurface ${\cal S}$ nowhere tangent to $V$. Pick a 1-form
$e^-$ that satisfies
\be
\label{eqn:eminus}
e^- \cdot V = 1 \ , \quad  (e^-)^2 = 0
\ee
on ${\cal S}$. Now propagate $e^-$ off ${\cal S}$ by solving ${\cal
L}_V e^- = 0$. Equations \reef{eqn:eminus} continue to hold because
$V$ is Killing. Let $e^+ = V$. Since $e^+$ and $e^-$ commute they must
be tangent to two dimensional surfaces in spacetime. These $2$-surfaces
form a $4$-parameter family $\Sigma_2 (x^m)$ where $m=1 \ldots 4$.
Since $e^+$ is a null Killing vector field, we know that it must be tangent to affinely
parametrized null geodesics. We can define a coordinate $v$ to be the
affine parameter distance along these geodesics. Choose another
coordinate $u$ so that $(u,v)$ are coordinates on the surfaces
$\Sigma_2$. Then
\be
 e^+ = \frac{\partial}{\partial v},
\ee
\be
 e^- = H \left( \frac{\partial}{\partial u} - \frac{\cal F}{2}
 \frac{\partial}{\partial v} \right),
\ee
for some functions $H$ and ${\cal F}$. $H$ must be non-zero
because $e^+$ and $e^-$ are not parallel. $H$ and ${\cal F}$ must be
independent of $v$ because $e^+$ and $e^-$ commute. We shall assume
that $H>0$, which can always be arranged by $u \rightarrow -u$.
Other than these
restrictions, $H$ and ${\cal F}$ are arbitrary and can be chosen to
be anything convenient. However, we shall keep them arbitrary because
different gauges are convenient for different solutions. This freedom
in choosing $H$ and ${\cal F}$ means that our general solution will
contain a lot of gauge freedom.

Using $(e^+)^2 = (e^-)^2 = 0$ and $e^+ \cdot e^- = 1$, the metric on
the surfaces $\Sigma_2$ can be deduced to take the form
\be
 ds_2^2 = H^{-1} \left( {\cal F} du^2 + 2 du dv \right) \ .
\ee
We shall take $(u,v,x^m)$ as the coordinates on our six dimensional
spacetime. Once the functions $x^m$ labelling the 2-surfaces have been chosen,
the coordinates $u$ and $v$ are only defined up to transformations of the form
\be
 \label{eqn:transfm}
 u = u' + U(x), \qquad v = v' + V(u',x).
\ee
In these coordinates, the six dimensional metric can be written
\be
\label{eqn:lineelt}
 ds^2 = 2 H^{-1} \left( du + \beta_m dx^m \right) \left( dv + \omega_m dx^m +
 \frac{\cal F}{2} \left(du + \beta_m dx^m \right) \right) - H h_{mn}
 dx^m dx^n,
\ee
where the metric $h_{mn}$ will be referred to as the metric on the
``base space'' ${\cal B}$ and $\omega$ and $\beta$ will be regarded as
1-forms on ${\cal B}$. The functions ${\cal F}$ and $H$, the 1-forms
$\omega$ and $\beta$ and the metric $h_{mn}$ all depend on $u$ and $x$
but not $v$ (because $V$ is Killing). Note that the only information
we have used so far is that $V$ is a null Killing vector field.

In these coordinates we have
\bea
\label{eqn:basis}
e^+ &=& H^{-1} \big(du + \beta_m dx^m\big)
\nn
e^- &=& dv +\omega_m dx^m + \frac{{\cal F}H}{2} e^+
\eea
and we can complete $e^+$ and $e^-$ to a null basis by defining
\be
\label{eqn:basis2}
e^a = H^{1 \over 2} \he^a{}_m dx^m
\ee
where $\he^a$ is a vierbein for ${\cal B}$, which we shall choose to
be independent of $v$. Note that this basis need {\it not} be the same as
that used in equations \p{eqn:antisd} and \p{eqn:constantspinor}, so
these equations will not hold in general.

It is convenient to define anti-self dual 2-forms on ${\cal B}$ by
\be
 J^i = H^{-1} I^i,
\ee
because one then finds
\be
 (J^i)^m{}_p (J^j)^p{}_n = \epsilon^{ijk} (J^k)^m{}_n - \delta^ {ij} \delta^m_n,
\ee
where the indices $m,n\ldots$ have been raised with $h^{mn}$. Hence,
these 2-forms yield an almost hyper-K\"ahler structure on ${\cal B}$.

We should emphasize that our introduction of coordinates is purely
local, valid only in some open subset of spacetime. In particular, there is
no reason why the notion of a base space should be valid globally. In
general, the only globally well-defined objects are $V$ and $X^i$.

\subsubsection*{Conditions for supersymmetry}

We shall now express the necessary and sufficient conditions for
supersymmetry in these coordinates.
It is convenient to define a restricted exterior derivative
${\tilde{d}}$ acting on $p$-forms defined on ${\cal B}$ as follows;
suppose $\Phi \in \Lambda^p ({\cal B})$ with
\be
 \Phi = {1 \over p!} \Phi_{m_1 \dots m_p} (x,u) dx^{m_1} \wedge \ldots \wedge
 dx^{m_p},
\ee
then let
\be
 {\tilde{d}} \Phi \equiv {1 \over (p+1)!} (p+1) {\partial \over
 \partial x^{[q}} \Phi_{m_1 \dots m_p]} dx^q \wedge  dx^{m_1}
 \wedge \ldots \wedge dx^{m_p}.
\ee
Next, we define the operator $\cD$ acting on such $p$-forms as
\be
 \cD \Phi = {\tilde{d}} \Phi - \beta
\wedge {\dot{\Phi}}
\ee
where ${\dot{\Phi}}$ denotes the Lie derivative of $\Phi$ with
respect to $\partial \over \partial u$.
Note that
\be d \Phi = \cD \Phi +H e^+ \wedge {\dot{\Phi}} \ee and
\be \cD^2 \Phi = - \cD \beta \wedge{\dot{\Phi}} \ .
\ee

With this choice of notation, we note that
\bea
\label{eqn:deplm}
de^+ &=& H^{-1} \cD \beta + e^+ \wedge (H^{-1} \cD H +  {\dot{\beta}})
\nn
de^- &=& \cD \omega +{\cF \over 2} \cD \beta +H e^+ \wedge
\big( \dot{\omega} +{\cF \over 2} {\dot{\beta}}
- {1 \over 2} \cD \cF \big).
\eea
Using these expressions, it is straightforward to compute
the components of the spin connection. In particular, $\cL_V
e^\alpha=0$ and \p{eqn:spafix} imply
\be
\label{eqn:minuscompts}
\omega_{\alpha \beta -}=- \omega_{- \alpha \beta} = {1 \over 2} (de^+)_{\alpha \beta}
\ee
and the remaining components are given in Appendix \ref{sec:spinconn}.
Equation \p{eqn:spafix} implies that
\be
 G_{-+a}e^a = {1 \over 2}  \big(H^{-1} \cD H+  {\dot{\beta}} \big),
\ee
and
\be
 {1 \over 2} G_{-ab} e^a \wedge e^b = {1 \over 2} H^{-1} \cD \beta.
\ee
The self-duality of $G$ now implies that
\be
{1 \over 6} G_{abc} e^a \wedge e^b \wedge e^c = {1 \over 2} \star_4 (\cD H + H  {\dot{\beta}}),
\ee
and
\be
\label{eqn:betasd}
 \cD \beta = \star_4 \cD \beta,
\ee
where $\star_4$ denotes the Hodge dual defined on ${\cal B}$. Hence
$\cD \beta$ is self-dual on ${\cal B}$. This implies that
equation \p{eqn:evensmp} holds for $\alpha=-$.

The remaining components of $G$ are $G_{+ab}$. These are obtained using the
$\alpha=+$ component of \p{eqn:evensmp}. It is straightforward to show that
\be
{1 \over 2} G_{+ab} e^a \wedge e^b = H \psi - {1 \over 2} (\cD \omega)^-
\ee
where ${}^\pm$ denotes the self-dual (anti-self-dual) projection on ${\cal B}$,
and
\be
\label{eqn:psidef}
\psi = {H \over 16} \epsilon^{ijk} (J^i)^{pq} ({\dot{J^j}})_{pq} J^k.
\ee
Using a coordinate transformation of the form $x \rightarrow x(u,x')$
it would be possible to reach a gauge in which $\psi = 0$ but we shall
keep things general here.

To summarize, \reef{eqn:spafix} and the $\alpha=+$ component of \p{eqn:evensmp}
together with the self-duality of $G$, fix $G$ to be
\bea
\label{eqn:Gexpand}
G &=& {1 \over 2} \star_4 \left(\cD H + H  {\dot{\beta}} \right)
+e^+ \wedge \left(H \psi - {1 \over 2} (\cD \omega)^-\right)
\nn
&+& {1 \over 2} H^{-1} e^- \wedge \cD \beta - {1 \over 2} e^+ \wedge e^-
\wedge \left(H^{-1} \cD H+ {\dot{\beta}} \right).
\eea
The remaining differential constraints are the $\alpha=c$ components of \p{eqn:evensmp}
which constrain the covariant derivatives of $J$ on ${\cal B}$. In fact, it suffices
to note that from the closure of $X^i$, we obtain
\be
\label{eqn:cdJ}
{\tilde{d}} J^i = \partial_u \big(\beta \wedge J^i \big)
\ee
where $\partial_u$ denotes the Lie derivative with respect to
$\partial/\partial u$. This, together with the fact that the $J^i$ satisfy the
quaternionic algebra, implies the  $\alpha=c$ components of \p{eqn:antisd}
(see for instance section 2 of \cite{GMW}). Equation \p{eqn:cdJ} shows
that the almost hyper-K\"ahler structure of ${\cal B}$ is not
integrable in general.

We have now exhausted the content of the algebraic and
differential constraints satisfied by $V$ and $I^i$ hence, as
explained above, the existence of a Killing spinor is
guaranteed. Therefore, in these coordinates, the necessary and sufficient
conditions for the existence of a Killing spinor are that the field
strength be given by \p{eqn:Gexpand}, that $\beta$ obey the
self-duality condition \p{eqn:betasd}, and that the complex structures
obey \p{eqn:cdJ}. We are interested in obtaining supersymmetric {\it
solutions} so we now turn to the field equations.

\subsubsection*{The Bianchi identity}

Having obtained an expression for $G$, we
need to solve the Bianch identity $dG=0$ (which is also the equation
of motion for $G$ because $G$ is self-dual). Using ({\ref{eqn:deplm}})
the Bianchi identity reduces to
\be
 \label{eqn:biana} \cD \left(\star_4 (\cD H + H
 {\dot{\beta}}) \right) + \cD \beta \wedge {\cal G}^+ =0 \ ,
\ee
and
\be
\label{eqn:bianb}
 \tilde{d} \left( {\cal G}^+ + 2 \psi \right) = \partial_u \left[ \beta
 \wedge \left( {\cal G}^+ + 2\psi \right) + \star_4 \left( \cD H + H \dot{\beta} \right)
 \right],
\ee
where we have introduced a self-dual 2-form
\be
 {\cal G}^+ \equiv H^{-1} \left( (\cD \omega)^+ + \frac{1}{2} {\cal F}
 \cD \beta \right).
\ee

\subsubsection*{The Einstein equation}

It remains to consider the Einstein equations. In fact, as we show in Appendix B,
the only component of the Einstein equations not implied by the Killing spinor
and gauge equations is the $++$ component. We must therefore compute
$R_{++}$ using the spin connection components given in Appendix \ref{sec:spinconn}.
It is useful to define
\be
L =   {\dot{\omega}} + \frac{1}{2} \cF {\dot{\beta}} -{1 \over 2} \cD \cF
\ee
so that $\omega_{++a}=-H L_a$. Then we obtain
\bea
R_{++} &=& \star_4 \cD (\star_4 L) + 2{\dot{\beta}}_m L^m
+{1 \over 2} H^{-2} \left( \cD \omega +{1 \over 2} \cF \cD \beta \right)^2
\nn
&-&{H \over 2} h^{mn} \partial_u^2 (H h_{mn})
-{1 \over 4}  \partial_u (Hh^{mn}) \partial_u (Hh_{mn})\ ,
\eea
where for $\Phi \in \Lambda^2 ({\cal{B}})$, $\Phi^2 \equiv (1/2) \Phi_{mn} \Phi^{mn}$.
Hence the Einstein equation reduces to
\bea
\label{eqn:einstn}
 \star_4 \cD (\star_4 L) &=&
 \frac{1}{2} H h^{mn} \partial_u^2 (H h_{mn})
+{1 \over 4}  \partial_u (Hh^{mn}) \partial_u (Hh_{mn})
- 2{\dot{\beta}}_m L^m
\nn
&+& \frac{1}{2} H^{-2}\left((\cD \omega)^- - 2 H \psi \right)^2
-{1 \over 2} H^{-2} \left( \cD \omega + \frac{1}{2} \cF \cD \beta \right)^2.
\eea

\subsubsection*{Summary}

We have obtained a general local form for all supersymmetric solutions of
minimal $D=6$ supergravity. The metric is given by \p{eqn:lineelt}
and the necessary and sufficient conditions for supersymmetry can be
expressed as a set of equations on the base manifold ${\cal B}$. This
must admit an almost hyper-K\"ahler structure with almost complex
structures obeying equation \p{eqn:cdJ}. The 1-form $\beta$ must obey the
self-duality condition \p{eqn:betasd}. In terms of the basis
\p{eqn:basis}, \p{eqn:basis2}, the field strength $G$ is given by
\p{eqn:Gexpand}. Finally, the Bianchi identity and Einstein equation
must be satisfied, which gives equations \p{eqn:biana}, \p{eqn:bianb}
and \p{eqn:einstn}.

\sect{The $G$-structure}

\label{Gstructures}

We have shown that any solution to the supersymmetry equation (\ref{susyeq})
is characterized by the existence of a set of forms which obey algebraic
and differential constraints. This fact is related to the notion of
$G$-structures.
The relevance of $G$ structures for classifying supersymmetric geometries
in supergravity theories was put forward in
\cite{GMPW} (see also \cite{Friedrich:2001nh}) 
and subsequently used to analyse and classify
supersymmetric solutions in various supergravity theories
in \cite{d5,dan,dallagata,GP,KMPT,GMW,KMT,GG,MS}.

A $G$-structure is a global reduction of the frame bundle, whose structure group
is generically $GL(n,\bR)$, to a sub-bundle with structure group $G$. This
reduction is equivalent to the existence of certain tensors whose isotropy
group is $G$. When these tensors are globally defined over the manifold, then
their isotropy group
is promoted to the structure group of the bundle. Here we assume that
six-dimensional space-time is equipped with a Lorentzian metric $g$ and
a spin structure, hence it has generically a $Spin(1,5)$ structure.
The existence of a globally defined spinor with isotropy group
$G \subset Spin(1,5)$ defines the  $G$-structure of relevance here.
Equivalently, this is defined by the spinorial bi-linears we have discussed
above.

According to \cite{bryant} there are four different types of stabilizer groups
for a spinor in $Spin(1,5)$. The one relevant here is that associated to
a chiral spinor and turns out to be the group
$SU(2)\ltimes \bb{R}^4 \subset Spin(1,5)$ (see also \cite{FOF}). 
Notice that in contrast to five dimensions \cite{d5} we have here only
one possible isotropy group of the spinor: since the corresponding  
Killing vector is everywhere null (and non-zero),
we have a  globally defined $SU(2)\ltimes \bR^4$ structure on spacetime.
Indeed the $D=5$ ``timelike'' and ``null'' cases discussed in \cite{d5}
correspond to the $SU(2)$ and $\bR^3$ subgroups of $SU(2)\ltimes \bR^4$
respectively. Note that these $D=5$ structures are only defined 
locally since a timelike Killing vector may become null somewhere. However,
they admit a unified global description in six dimensions. In section \ref{dimred}
we will describe explicitly the reduction from six to five dimensions. 

In terms of the Killing spinor, one can give an explicit demonstration of the
$SU(2)\ltimes \bR^4$ structure exploiting the isomorphism $Spin(1,5)\simeq SL
(2, \bH)$
 \cite{bryant}, namely one can realize $Spin(1,5)$ as 2 $\times$ 2
 quaternionic-valued matrices $A$ with unit determinant. They act on the space
 of spinors identified with $\bH^2\oplus \bH^2$ as $A \cdot (s_+, s_-)=
 (As_+,(A^*)^{-1}s_-)$, with $s_{+/-}$  corresponding to the
 positive/negative chirality spinor. A chiral spinor $s_+$ therefore has
 stabilizer group
\bea
s_+ = \left(
\begin{array}{c}
 r\\
 0 \\
\end{array}
\right)~, \qquad \qquad
G = \left\{
\left(
\begin{array}{cc}
 1 & a \\
 0 & b
\end{array} \right)| a \in \bH, ~b \in SU(2)\right\}\simeq SU(2)\ltimes \bR^4~.
\eea
An alternative way to derive this is, following \cite{FOF},   to write explicit
representations  for the Clifford algebra $Cl (1,5)\simeq Cl (0,4)\otimes
Cl (1,1)$ and the corresponding spinor on which they act. From this it is not
difficult to show that the  algebra which leaves the spinor invariant has
generators
\be
\frac{1}{2} a_{ab}\gamma_{ab} + b_c \gamma_{-c}
\ee
where $b \in \bR^4$ and $a_{ab}$ are such that  $a_{ab}\gamma_{ab}$ fixes the
spinor in $Cl (0,4)$, i.e. they span the $su(2)$ algebra.

The existence of the $SU(2)\ltimes \bR^4$ structure implies that one can 
introduce a local null frame $\{e^+,e^-,e^a\}$ in which the metric, 
one-form and three-forms are written as 
\be
ds^2 = 2 e^+ e^- - \delta_{ab} e^a  e^b
\ee
and
\be
V =   e^+, \qquad X^i =   e^+ \wedge I^i~,
\ee
where $I^i$ obey the algebra of the quaternions. One can indeed check 
that these are invariant under the $SU(2)\ltimes \bR^4$ action on the 
tangent space  given by
\bea
e^{+'} & = & e^+\\
e^{-'} & = & e^- + q^a q^a e^+ + \sqrt{2} q^a M^a_b e^b\\
e^{a'} & = & M^a_b e^b + \sqrt{2} e^+ q^a
\eea
for any $q^a \in \bR^4$ and $M^a_b \in SO(3)$.

The type of $G$-structure is determined  completely by covariant
derivatives of the spinor, or of the forms,  and is characterized in terms of
its intrinsic torsion which lies in the space
 $\Lambda^1\otimes g^\perp$ and
decomposes under irreducible $G$-modules. Thus if all of the components
vanish the Levi-Civita connection has holonomy contained in $G$. Manifolds
with $SU(2)\ltimes \bR^4$ are discussed in \cite{FOF,bryant}.
In general, when we have a non-trivial $G$ field turned on, the holonomy
of the Levi-Civita connection is not in $SU(2)\ltimes \bR^4$, and
departure from special holonomy is measured by  the intrinsic torsion.

In our context, a convenient way to express the constraints on the
intrinsic torsion is in terms
of geometrical data on the base manifold ${\cal B}$. Although this is clearly
not globally defined, in each patch the local form of the metric is given 
by  \reef{eqn:lineelt} and instead of the globally defined objects 
$\{V,X^i\}$
 one can equivalently 
express the differential conditions in terms of $\{\beta,J^i\}$. These encode 
information about the ($u$-dependent) almost hyper--K\"ahler structure 
of ${\cal B}$. We have shown that
supersymmetry and self-duality of the $G$ field are equivalent
to the following constraints on the base manifold
\bea
\tilde{d} J^i  &= & \partial_u \left(\beta \wedge J^i \right)\nn
\cD \beta & =  & \star_4 \cD \beta~.
\label{base:eqs}
\eea
Once these two (coupled) conditions are fulfilled, then the $G$ field
is explicitly determined by equation
\reef{eqn:Gexpand}.
Note that in general the almost hyper--K\"ahler structure
is completely generic in terms of the  three
irreducible components of its intrinsic torsion
$(\mathbf{2} + \bar\mathbf{2} ) + (\mathbf{2} +
\bar\mathbf{2} ) +(\mathbf{2} + \bar\mathbf{2} ) $
(see e.g. \cite{GMW}), thus for
instance ${\cal B}$ is not a complex or a K\"ahler manifold. This is certainly
an unpleasant complication when it comes to seeking general examples. In the remainder
of the paper we will discuss in detail some interesting and rather general cases
where we do have some control over the base space. Although
\reef{base:eqs} (together with \p{eqn:Gexpand}) are sufficient
to ensure supersymmetry, we recall that we must also
impose the Bianchi identity and the Einstein equation to get solutions of the
supergravity theory.

\sect{Special cases}

\label{sec:examples}

\subsection{Non-twisting solutions}

If $V \wedge dV$ vanishes everywhere then the congruence of null
geodesics tangent to $V$ has vanishing twist. Such solutions will
therefore be referred to as {\it non-twisting}. For non-twisting
solutions, $V$ is hypersurface orthogonal, and hence there exist
functions $H$ and $u$ such that
\be
 V = H^{-1} du.
\ee
Therefore, in the non-twisting case, there is a preferred definition
of $H$ and $u$, in contrast with the general (twisting)
case where the definition is gauge-dependent. Comparing with equation
\reef{eqn:basis} we see that non-twisting solutions have
\be
 \beta = 0.
\ee
Some gauge freedom remains in the definition of the coordinates $v$
and $x^m$: one is still free to perform coordinate transformations of
the form $v \rightarrow v - V(u,x)$ and $x^m \rightarrow
x^m(u,x')$. This freedom could be used to set $\omega = 0$ or $\psi=0$, for
example, but we shall keep $\omega$ and $\psi$ general here. The metric of a
non-twisting solution takes the form of a plane-fronted wave:
\be
 ds^2 = 2 H^{-1} du \left( dv + \omega_m dx^m +
 \frac{\cal F}{2} du \right) - H h_{mn}
 dx^m dx^n.
\ee
Then ({\ref{eqn:cdJ}}) implies that
\be
 \tilde{d} J^{i}=0
\ee
so the $J^i$ define an integrable hyper-K\"ahler structure on ${\cal
B}$, i.e., ${\cal B}$ is hyper-K\"ahler. From ({\ref{eqn:Gexpand}}) we obtain
\bea
\label{eqn:Gexpandb}
G = {1 \over 2} \star_4 ({\tilde{d}} H)
+e^+ \wedge \left( H \psi - {1 \over 2} ({\tilde{d}} \omega)^- \right)
-{1 \over 2} H^{-1} e^+ \wedge e^- \wedge {\tilde{d}} H \ .
\eea
There is also considerable simplification to the Bianchi and Einstein equations.
In particular, from ({\ref{eqn:biana}}) we find
\be
 \tilde{\nabla}^2 H = 0,
\ee
where $\tilde{\nabla}$ is the Levi-Civita connection of ${\cal B}$.
Hence $H$ is harmonic on ${\cal B}$. Equation ({\ref{eqn:bianb}}) simplifies to
\be \label{eqn:bianbb}
 {\tilde{d}} \left(H^{-1}   ({\tilde{d}} \omega)^+ + 2\psi \right)
 = \partial_u  \left( \star_4 ({\tilde{d}} H) \right)
\ee
and the Einstein equation \p{eqn:einstn} becomes
\bea
\label{eqn:einstnb}
 {\tilde{\nabla}}^m (\dot{\omega})_m  -{1 \over 2} {\tilde{\nabla}}^2 \cF  &=&
-{1 \over 2} H h^{mn} \partial_u^2 (H h_{mn})
-{1 \over 4}  \partial_u (Hh^{mn}) \partial_u (Hh_{mn})
\nn
&-& \frac{1}{2} H^{-2} \left(( {\tilde{d}} \omega)^- - 2 H \psi \right)^2
+ {1 \over 2} H^{-2} \left( {\tilde{d}} \omega  \right)^2.
\eea
Note that these equations can be solved successively:
first one picks a ($u$-dependent) hyper-K\"ahler base space ${\cal B}$,
then a ($u$-dependent) harmonic function $H$ on ${\cal B}$, then one
seeks a $1$-form $\omega$ that solves equation \reef{eqn:bianbb} and
finally a function ${\cal F}$ satifying equation \reef{eqn:einstnb}.

\subsubsection*{Flat base space}

To construct examples of non-twisting solutions with a flat base space,
we take the base space to be flat $\bR^4$ with
the metric written in terms of either left-invariant $\sigma_R^i $ or
right-invariant  $\sigma^i_L$ one-forms on the three-sphere:
\bea
   d s^2 = d r^2
      + \frac{1}{4}r^2\left(
         (\sigma^1)^2 + (\sigma^2)^2+ (\sigma^3)^2 \right)
\eea
where $d \sigma^i = {1 \over 2} \eta \epsilon^{ijk} \sigma^j \wedge \sigma^k$,
and $\eta=1$ if $\sigma=\sigma_R$, $\eta=-1$ if $\sigma=\sigma_L$. We take
an orthonormal basis on $\bR^4$ given by
\be
e^1 = dr \ , \quad e^2 = {r \over 2} \sigma^1 \ , \quad
e^3 = {r \over 2} \sigma^2 \ , \quad e^4 = {r \over 2} \sigma^3
\ee
with positive orientation defined with respect
to $e^1 \wedge e^2 \wedge e^3 \wedge e^4$.
We take hyper-K\"ahler structures $J^i$ defined via
$J^i = -{\eta \over 4} r^{2(1+\eta)} d (r^{-2 \eta} \sigma^i)$; hence $\psi=0$.
We assume that the harmonic function $H$ depends only on $u$ and $r$, so
\be
H = P(u) + \frac{Q(u)}{r^2}
\ee
for arbitrary functions $P$, $Q$ to be fixed. It remains to solve for
the remainder of the Bianchi identity and the Einstein equation. These simplify to
\be \label{eqn:bianbbc}
 {\tilde{d}} \left( H^{-1}  ({\tilde{d}} \omega)^+ \right)
 =  \partial_u \left(\star_4 ({\tilde{d}} H) \right)
\ee
and
\be
\label{eqn:einstnbcd}
{\tilde{\nabla}}^m (\dot{\omega})_m- {1 \over 2} {\tilde{\nabla}}^2 \cF   =
-2H \ddot{H} - \dot{H}^2 +{1 \over 2} H^{-2} \left( ( {\tilde{d}}
\omega )^+  \right)^2.
\ee
To find a solution to these equations, we assume that
\be
\omega = W(u,r) \sigma^3
\ee
and $\cF=\cF(u,r)$. Substituting into ({\ref{eqn:bianbbc}}) we obtain
\be
 \dot{Q}=0,
\ee
so $Q$ is constant, together with
\be
W= {\alpha_2}(u) r^{-2 \eta}+{1 \over 2} {\alpha_1}(u)
\left( {P \over 2 \eta} r^{2 \eta}+{Q \over 2 \eta-1}r^{2 \eta-2} \right)
\ee
for arbitrary functions ${\alpha_1}(u)$,  ${\alpha_2}(u)$. Lastly, we solve
({\ref{eqn:einstnbcd}}) for $\cF$. We obtain
\bea
\label{eqn:flatfsol}
\cF = {\alpha_3}(u)+{\alpha_4}(u) r^{-2} +{1 \over 2} \left(P
\ddot{P}+{1 \over 2}\dot{P}^2 \right) r^2- {{\alpha_1}(u)^2 \over 4
  \eta (2 \eta -1)}r^{4 \eta -2}+2Q \ddot{P} \log r
\eea
for arbitrary functions ${\alpha_3}(u)$,  ${\alpha_4}(u)$.
Observe that $P$ must be linear in $u$ in order for the logarithmic
term in ({\ref{eqn:flatfsol}}) to vanish.

It is clear that this treatment can be extended to other examples of hyper-K\"ahler base space,
for example Eguchi-Hanson or Taub-NUT space with $u$-dependent parameters; and a large family
of new solutions can be constructed in this manner. We shall however not pursue this here.

\subsubsection*{pp-waves}

Our general non-twisting solution describes a pp-wave if $du$ is
covariantly constant, which happens if, and only if,
$H = {\rm constant}$. By rescaling the coordinates we can take
$H \equiv 1$ so the solution becomes
\bea
 ds^2 & = & 2 du \left( dv + \omega_m dx^m +
 \frac{\cal F}{2} du \right) - h_{mn}dx^m dx^n~,\nn
 G  & = & du \wedge (\psi -\frac{1}{2}(\tilde{d}\omega)^-)~
\eea
with $h_{mn}$ a hyper-K\"ahler metric. As mentioned above, we can
always change coordinates $x \rightarrow x(u,x')$ so that $\psi=0$ in
the new coordinates. Alternatively, the same type of coordinate
transformation could be used to make $\omega$ vanish. However, in
general it is not possible to find a gauge in which both $\psi$ and
$\omega$ vanish. As an illustration of this point we will derive the maximally
supersymmetric plane wave solution in two ways: first in the gauge
$\psi=0$, and then in the gauge $\omega=0$, using a flat base space in both
cases.

When $\psi=0$ we can just consider a special case of the flat base
solution derived  above. In particular, set $P=1$, $\alpha_2 = 1/2$,
$\alpha_1=\alpha_3=\alpha_4=Q=0$ and $\eta=-1$. Converting to
Cartesian coordinates on $\bR^4$ (see e.g. \cite{d5}) this is
\be
\omega = \frac{1}{2} r^2 \sigma^3_L = x^1 dx^2-x^2dx^1 -x^3
dx^4+x^4 dx^3~.
\ee
Performing the following change of variables
\bea
x^1 & = & \cos u\, y^1 - \sin u \, y^2\nn
x^2 & = & \sin u\, y^1 + \cos u \, y^2\nn
x^3 & = & \cos u\, y^3 + \sin u \, y^4\nn
x^4 & = & -\sin u\, y^3 + \cos u \, y^4
\label{changeofvar}
\eea
we obtain the maximally supersymmetric plane wave as given in \cite{meesnote}
\bea
\label{eqn:kgsol}
ds^2 &=& 2 du\left(dv + \frac{1}{2}y^iy^i du\right)- d\mathbf{y}^2\nn
G & = & -du\wedge (dy^1 \wedge dy^2 - dy^3 \wedge dy^4 )~.
\label{KG6solution}
\eea
Let us now derive this solution directly in the gauge $\omega=0$.
On flat space we have the standard complex structures
\be
 K^{1} = dx^1 \wedge dx^2 - dx^3 \wedge dx^4, \qquad K^{2} = dx^1 \wedge dx^3 + dx^2
 \wedge dx^4, \qquad K^3 = dx^1\wedge dx^4 - dx^2\wedge dx^3.
\ee
We clearly cannot obtain the above solution by taking $J^i=K^i$ so
instead we shall take a triplet of $u$-dependent complex structures defined by
\bea
J^1 & = & K^1\nn
J^2 & = & \cos 2u \,K^2 + \sin 2u \,K^3\\
J^3 & = & -\sin 2u \,K^2 + \cos 2u \,K^3~.\nonumber
\eea
Thus we have
\be
\psi = - K^1 \ .
\ee
With this choice, the Bianchi identity holds automatically, and from the Einstein equation we obtain
\be
{\tilde{\nabla}}^2 \cF  = 4\psi^2 = 8
\ee
which is solved by taking $\cF = x^i x^i$. The solution is then the
same as \p{eqn:kgsol} with $y^i=x^i$.

\subsection{$u$-independent solutions}

Another instance in which the general equations simplify
considerably is when there is no dependence of the solution on the co-ordinate
$u$. Geometrically, we can characterize this case by the existence of
a second Killing vector field $K$ which commutes with $V$, and is
not orthogonal to $V$. $V$ and $K$ are then tangent to timelike
2-surfaces so we can use these as the 2-surfaces $\Sigma_2$ in our
introduction of coordinates, with $K=\partial/\partial u$. If we also
assume that $K$ preserves the 3-forms $X^i$ then we can drop all
$u$-dependence in our equations.

For such solutions, the base space ${\cal B}$ is hyper-K\"ahler since
\be
 \tilde{d} J^{i} = 0~,
\ee
and $\beta$ has self-dual curvature on ${\cal B}$
\be
\label{eqn:betasd2}
 \tilde{d} \beta = \star_4 \tilde{d} \beta~.
\ee
The Bianchi identity and Einstein equation reduce respectively to
\be
\label{eqn:uindep1}
 \tilde{d} \star_4 \tilde{d} H + \tilde{d} \beta \wedge {\cal G}^+  =  0~,
\ee
\be
\label{eqn:uindep2}
 \tilde{d} {\cal G}^+ =0~ ,
\ee
and
\be
\label{eqn:uindep3}
 \tilde{\nabla}^2 {\cal F} = -\left( {\cal G}^+ \right)^2 \ ,
\ee
where ${\cal G}^+$ is given by
\be
 {\cal G}^+ = H^{-1} \left(
   (\tilde{d}\omega)^+ + \frac{1}{2} {\cal F} \tilde{d} \beta \right)~.
\ee

\subsubsection*{An example}

As an example of such a solution, take $H=1$, ${\cal F} = 0$ and the
base space to be flat with $u$-independent complex structures and line element
\be
 ds^2 = dx^i dx^i,
\ee
and
\be
 \beta ={1 \over \sqrt{2}}( x^1 dx^2 - x^2 dx^1 + x^3 dx^4 - x^4 dx^3),
\ee
\be
 \omega ={1 \over \sqrt{2}}( x^1 dx^2 - x^2 dx^1 - x^3 dx^4 + x^4 dx^3).
\ee
Constants multiplying $\beta$ and $\omega$ can be absorbed into an
overall scale by rescaling the coordinates. 
Let $u = {1 \over \sqrt{2}}(t+z)$ and $v={1 \over \sqrt{2}}(t-z)$ to obtain the
metric
\be
 ds^2 = (dt + x^1 dx^2 - x^2 dx^1)^2 - (dx^1)^2 - (dx^2)^2 - (dz+x^3
 dx^4 - x^4 dx^3)^2 - (dx^3)^2 - (dx^4)^2.
\ee
The field strength is
\be
 G = (dt + x^1 dx^2 - x^2 dx^1)\wedge dx^3 \wedge dx^4 - (dz + x^3 dx^4
 - x^4 dx^3) \wedge dx^1 \wedge dx^2.
\ee
The metric is a direct product of the metric of a three dimensional
G\"odel universe (first constructed in \cite{horowitz:95}) with a
three dimensional internal space. However, the field strength is not
a direct product. The internal space is homogeneous with isometry
group $Nil$. The G\"odel universe is also homogeneous.

It has been shown \cite{boyda:03,harmark:03} that supersymmetric G\"odel
universes can be related via T-duality to supersymmetric plane wave
solutions. For the solution above, this works as follows. First, we
write it as a solution of type II supergravity with constant dilaton
and self-dual three form flux $H = 2 G$. In order to perform a
T-duality along the $z$ direction it is convenient to
choose the following gauge for the $B$ field
\be
B = \left( dz + x^3 dx^4 -x^4 dx^3 \right) \wedge \left( dt + x^1 dx^2
-x^2 dx^1 \right).
\ee
After T-duality we obtain the following metric and NS-NS field strength
\bea
ds^2 &= & 2dz \left( dt + (x^1 dx^2 - x^2 dx^1)- \frac{1}{2}dz\right) -
 d{\bf x}^2 - d{\bf z}^2\nn
 H & = & -2 dz \wedge dx^3 \wedge dx^4~,
\eea
where $z^i$ are the $4$ flat directions arising in the oxidation to
$D=10$. Note that $H$ is not self-dual, hence this does not give a solution of
miminal $D=6$ supergravity. Performing a change of variables
similar to \reef{changeofvar} in the $(x^1,x^2)-$plane the solution reads
\bea
ds^2 & = & 2du'\left(dv' + \frac{1}{2} (y_1^2 + y_2^2)du'\right)-
d{\bf y}^2 -d{\bf z}^2 \nn
H & = & - 2 du' \wedge dy^3 \wedge dy^4
\eea
where we have defined $u'=z, v'=t-\frac{1}{2}z$. Thus the metric looks like
$CW_4\times \bR^6$ and is a homogeneous plane wave. The field $H$ however breaks
the symmetry of the solution because it has mixed components. This in
particular shows that the solution is not a parallelizable plane wave
\cite{sadri,fof:parallel}.

\subsection{Dimensional reduction}
\label{dimred}

Kaluza-Klein reduction of minimal six dimensional supergravity on a
circle yields a five dimensional supergravity theory. The reduction
yields 1 KK vector from the metric, 1 from the 2-form potential and 1
from dualizing the 3-form field strength. However, self-duality of
this 3-form implies that only 2 of these vectors are independent. One
also obtains a dilaton from the reduction of the metric. Hence the
$D=5$ theory consists of minimal $D=5$ supergravity coupled to a $D=5$
vector multiplet.

It is of interest to examine how the supersymmetric solutions of
{\it minimal} five dimensional supergravity obtained in \cite{d5} arise
from the six dimensional theory (this has already been done for some
maximally supersymmetric solutions \cite{ortin}). The details of the dimensional
reduction are given in \cite{ortin,herdeiro:02}. It is convenient to
consider the five dimensional timelike and null classes separately.

\subsubsection*{Timelike solutions}

The $D=5$ timelike class can be
obtained by dimensional reduction of a subset of our $u$-independent
(generically twisting) solutions as follows.
Solutions with no $u$-dependence can be
Kaluza-Klein reduced to $D=5$ provided $\partial/\partial u$
is spacelike, i.e., provided ${\cal F}$ is negative. The $D=6$ line element can
be written
\be
 ds_6^2 = H^{-1} {\cal F} \left[ du + \beta + {\cal F}^{-1}
 (dv + \omega) \right]^2 - H^{-1} {\cal F}^{-1} (dv + \omega)^2 - H
 ds_4^2.
\ee
The minimal $D=5$ theory does not contain a dilaton so we take
${\cal F} = - H$. Consistency of equations \reef{eqn:uindep1} and
\reef{eqn:uindep3} then requires $\tilde{d}\beta = {\cal G}^+$ hence
\be
\label{eqn:trunc}
 \tilde{d} \beta = \frac{2}{3} H^{-1} (\tilde{d} \omega)^+.
\ee
Now introduce some notation: let $t=v$, $f = H^{-1}$ and $G^+ =
f(\tilde{d}\omega)^+$ so ${\cal G}^+ = (2/3) G^+$. The five dimensional metric
is therefore
\be
 ds_5^2 = f^2 (dt + \omega)^2 - f^{-1} ds_4^2,
\ee
and equations \reef{eqn:uindep1} and \reef{eqn:uindep2} can be rewritten
as
\be
 \tilde{\nabla}^2 f^{-1} = \frac{4}{9} (G^+)^2
\ee
and
\be
 \tilde{d} G^+ = 0.
\ee
This reproduces the timelike class of supersymmetric solutions of minimal five
dimensional supergravity as given in equations \p{eqn:d5metric}, \p{eqn:d5eqs}. It can be verified
that the reduction of the $D=6$ field strength correctly reproduces the
$D=5$ field strength. Hence the $D=5$ timelike solutions are obtained
by taking ${\cal F}=- H$ and choosing $\beta$ to satisfy equation
\reef{eqn:trunc} (which is just a
consistency condition for the truncation required in reducing the
$D=6$ theory to the minimal $D=5$ theory). Note that $D=5$
solutions with $G^+ = 0$ arise from six dimensional solutions with
$\beta = 0$, i.e., $u$-independent non-twisting solutions.

\subsubsection*{Null solutions}

 The $D=5$ and $D=6$ metrics are related
by \cite{herdeiro:02}
\be
 ds_6^2 = ds_5^2 - \left( dz - \frac{2}{\sqrt{3}} A \right)^2,
\ee
where $F=dA$ is the five dimensional Maxwell field strength and $z$
the coordinate around the Kaluza-Klein circle. The five
dimensional null solution is \cite{d5}
\ba
 ds_5^2 &=& H^{-1} \left( {\cal F}_5 du^2 + 2 du dv \right) -
 H^2 \left(d{\bf x} + {\bf a}du \right)^2, \nn
 F &=& -\frac{H^{-2}}{2\sqrt{3}} \epsilon_{ijk} \nabla_j (H^3 a_k) du
 \wedge dx^i - \frac{\sqrt{3}}{4} \epsilon_{ijk} \nabla_k H dx^i
 \wedge dx^j,
\ea
where bold letters denote quantities transforming as 3-vectors and
$\nabla_i = \partial/\partial x^i$. $H$ is a $u$-dependent function
harmonic on $\bR^3$ that must also obey
\be
\label{eqn:integ2}
 \partial_u \nabla H = \frac{1}{3} \nabla \times \left[ H^{-2} \nabla
   \times (H^3 {\bf a}) \right].
\ee
The function ${\cal F}_5$ satisfies a Poisson-like equation
\cite{d5}. Solving for $A$ gives
\be
 A = A_u du - \frac{\sqrt{3}}{2} \chi_i dx^i,
\ee
where $\bchi$ satifies
\be
 \nabla \times \bchi = \nabla H,
\ee
which admits solutions because $H$ is harmonic. $A_u$ is obtained by
solving
\be
 \nabla A_u = - \frac{\sqrt{3}}{2} \partial_u \bchi +
 \frac{1}{2\sqrt{3}} H^{-2} \nabla \times (H^3 {\bf a}),
\ee
which admits solutions because the integrability condition
\reef{eqn:integ2} is satisfied. Using these results, the six
dimensional metric is
\be
 ds_6^2 = 2 H^{-1} du \left[ dv + \omega_i dx^i + \omega_z dz +
   \frac{1}{2} {\cal F} du \right] - H \left[H d{\bf x}^2 + H^{-1}
   \left(dz + \chi_i dx^i \right)^2 \right],
\ee
where
\be
 \omega_i = \frac{2}{\sqrt{3}} H A_u \chi_i - H^3 a_i,
\ee
\be
 \omega_z = \frac{2}{\sqrt{3}} H A_u,
\ee
\be
 {\cal F} = {\cal F}_5 - H^3 {\bf a}^2 - \frac{4}{3} H A_u^2.
\ee
The six dimensional solution belongs to our non-twisting
family of solutions. The base space is a Gibbons-Hawking
space \cite{gibbons:78} with harmonic function $H$. In general, this
will be $u$-dependent.

In summary, we have shown how all supersymmetric solutions of minimal
$D=5$ supergravity arise from supersymmetric solutions of minimal
$D=6$ supergravity. The $D=5$ timelike class arise from $D=6$
solutions for which $\partial/\partial u$ is a spacelike Killing
vector field whereas the $D=5$ null class arise from non-twisting $D=6$
solutions
for which the base space is a Gibbons-Hawking space (and therefore
admits a Killing vector field appropriate for dimensional
reduction). Some solutions can be reduced in both ways, for
example the $D=6$ maximally supersymmetric plane wave of \cite{meesnote} can
be reduced either to the (timelike) $D=5$ G\"odel solution of
\cite{d5} or the (null) $D=5$ maximally supersymmetric plane wave
(also given in \cite{meesnote}).

\subsection{Chiral null models}

Our non-twisting solutions closely resemble the ``chiral null
models'', a class of exact classical string backgrounds obtained in
\cite{horowitz:95}. (These generalize earlier classes of exact string
backgrounds obtained in \cite{horowitz:94a,horowitz:94b}. Another
family of solutions was obtained by duality in
\cite{lunin:02,lunin:03} but these solutions reduce to chiral null
models when restricted to minimal $D=6$ supergravity.) 
When $H$ and the base space are independent of $u$, our non-twisting solutions
{\it are} chiral null models, provided only that the choice of base
space corresponds to an exact ``transverse'' CFT. This is guaranteed
by the hyper-K\"ahler nature of our base space \cite{cvetic:96a}.
Hence, using the results of the previous subsection, all $D=5$ timelike
solutions with $G^+ = 0$ and all $D=5$ null solutions with
$u$-independent $H$ are exact classical string backgrounds. (Note that the
latter family includes the entire null class of minimal $N=2$, $D=4$
supergravity, which can be obtained by dimensional reduction \cite{d5}.)

In fact, there exist generalizations of chiral null models which describe exact
string backgrounds even when $H$ depends on $u$ \cite{callan:96}, and
examples of exact string backgrounds with a $u$-dependent transverse
space \cite{tseytlin:92}. It would be interesting to know how large a
class of exact string backgrounds is contained in our class of non-twisting
solutions.

Chiral null models are always non-twisting (i.e. the null Killing
vector field is hyper-surface orthogonal). However, an example of a
{\it twisting} exact string background was presented in
\cite{horowitz:95}. A large family of such ``generalized chiral null
models'' was considered in \cite{tseytlin:96}. It was suggested
(although not proved) that these are all exact string backgrounds,
again assuming an exact transverse CFT. Our $u$-independent solutions
are examples of such solutions. If these are indeed  
exact classical string backgrounds then it follows 
that the entire timelike class of the minimal $D=5$
theory must also be exact string backgrounds (and this in turn
includes the timelike class of minimal $N=2$, $D=4$ supergravity
\cite{d5}).

In general, supersymmetric solutions of minimal $D=6$ supergravity are
twisting {\it and} $u$-dependent. It would be interesting to know
which of these solutions describe exact classical string backgrounds.

\subsection{Solutions with Gibbons-Hawking base space}

The equations satisfied by our $u$-independent twisting solutions are
non-linear. It was argued in \cite{tseytlin:96} that such solutions
could not satisfy the superposition principle expected of BPS
objects. Here we shall show that this reasoning is incorrect by
considering solutions with a Gibbons-Hawking \cite{gibbons:78} base
space, i.e, the most general hyper-K\"ahler 4-manifold admitting a
Killing vector field $\partial/\partial z$ preserving the three
complex structures \cite{gibbons:88}:
\be
 ds_4^2 = H_2^{-1} \sigma^2 + H_2 d{\bf x}^2,
\ee
where
\be
 \sigma = dz + \chi_i dx^i,
\ee
$i=1,2,3$, $H_2$ and $\chi_i$ are independent of $z$, and
\be
 \nabla^2 H_2 = 0 \qquad, \nabla \cross \bchi = \nabla H_2,
\ee
where $\nabla_i \equiv \partial_i$ in this subsection.

We shall obtain all $u$-independent twisting solutions with a
Gibbons-Hawking base space for which the Gibbons-Hawking Killing
vector field $\partial/\partial z$ extends to a symmetry of the full
spacetime. This was done for the minimal $D=5$ theory in \cite{d5},
where it was shown that the solution is specified by four harmonic
functions of $x^i$. The analysis for the $D=6$ is very similar so we
shall just sketch the details here. Introduce an orthonormal basis on
the base space
\be
 e^0 = H_2^{-1/2} \sigma, \qquad e^i = H_2^{1/2} dx^i
\ee
so that the base space has orientation given by the volume form
$ e^0 \wedge e^1 \wedge e^2 \wedge e^3$. Let
\be
 \beta = \beta_0 \sigma + \beta_i dx^i, \qquad \omega = \omega_0
 \sigma + \omega_i dx^i.
\ee
The general solution of equation \reef{eqn:betasd2} is given by
\be
 \beta_0 = H_2^{-1} H_3, \qquad \nabla \times \bbeta = - \nabla H_3,
\ee
where $H_3$ is an arbitrary harmonic function of ${\bf x}$.
Self-duality implies that ${\cal G}^+$ must take the form
\be
 {\cal G}^+ = -\frac{1}{2} C_i e^0 \wedge e^i - \frac{1}{4}
 \epsilon_{ijk} C_k e^i \wedge e^j.
\ee
Solving equation \reef{eqn:uindep2} yields
\be
\label{eqn:csol}
 {\bf C} = 2 \nabla \left( H_2^{-1} H_4 \right),
\ee
where $H_4$ is an arbitrary harmonic function of ${\bf x}$.
Substituting these results into \reef{eqn:uindep1} and \reef{eqn:uindep3}
gives respectively
\be
 H = H_1 + H_2^{-1} H_3 H_4,
\ee
\be
 {\cal F} = - H_5 - H_2^{-1} H_4^2,
\ee
where $H_1$ and $H_5$ are further arbitrary harmonic functions of
${\bf x}$. Finally the definition of ${\cal G}^+$ yields an equation
for $\omega$:
\be
\label{eqn:bomegaeq}
 H_2 \nabla \omega_0 - \omega_0 \nabla H_2 - \nabla \times \bomega = 2
 \left(H_1 H_2 + H_3 H_4 \right) \nabla \left( H_2^{-1} H_4 \right) +
 \left( H_4^2 + H_2 H_5 \right) \nabla \left(H_2^{-1} H_3 \right).
\ee
Taking the divergence of this gives an integrability condition which
can be solved to determine $\omega_0$:
\be
\label{eqn:omega0}
 \omega_0 = H_2^{-2} H_3 H_4^2 + H_1 H_2^{-1} H_4 + \frac{1}{2}
 H_2^{-1} H_3 H_5 + H_6,
\ee
where $H_6$ is yet another arbitrary harmonic function of ${\bf
x}$. Substituting this back into \reef{eqn:bomegaeq} gives an equation
that determines $\bomega$ up to a gradient (which can be eliminated by
shifting $v$).

We have obtained the most general $u$-independent solution with a
Gibbons-Hawking base space whose Killing vector field extends to a
symmetry of the full solution. It is determined by $6$ arbitrary harmonic
functions of ${\bf x}$. It is clear that such solutions can be freely
superposed, as expected for solutions describing BPS states, in spite of
the non-linearity of the equations derived above.

If ${\cal F}<0$ then one can Kaluza-Klein reduce these solutions in
the $u$ and $z$ directions to
yield solutions of (non-minimal) $D=4$ supergravity. In general, this
reduction yields 2 KK vectors from the metric and 2 from the
2-form gauge potential, so the $D=4$ solution is parametrized by 8
charges: 4 electric and 4 magnetic \cite{cvetic:96b}. However, the
requirement of self-duality of the 3-form field strength reduces the
number of independent vectors to 3, so there are 6 charges,
corresponding to our 6 independent harmonic functions.
Taking (coincident) point sources for the harmonic functions generally leads
to $D=4$ solutions describing rotating naked singularities
(supersymmetric rotating black holes in $D=4$ apparently don't exist). However,
by demanding $\bomega=0$ one can obtain regular static supersymmetric black
holes. These are related by duality to the generating solutions
of \cite{cvetic:96c,bertolini:99}.

These solutions can also be reduced to $D=5$ to obtain solutions of
minimal $D=5$ supergravity coupled to a vector multiplet. KK reduction in the
$z$-direction yields $u$-independent null solutions in a similar
manner to subsection \ref{dimred}. Reduction in the $u$-direction
yields $D=5$ timelike solutions with Gibbons-Hawking base space,
generalizing those of \cite{d5} (to which they reduce when $H_5=H_1$
and $H_4=H_3$). This class of solutions includes supersymmetric
rotating black holes \cite{bmpv,tseytlin:96a,gauntlett:99} (assuming a
flat base space: $H_2=1/|{\bf x}|$).

If the $D=6$ solution is non-twisting then $\beta=0$ so one can set $H_3
=0$, which simplifies matters considerably. According to the the
discussion of the previous subsection, such solutions are chiral null
models and have been well-studied. For example, if one also sets
$\bomega=0$ then equation \reef{eqn:bomegaeq} imposes one further
condition, reducing the number of independent harmonic functions to
4. These solutions are the subclass of the solutions of
\cite{cvetic:96} corresponding to self-dual field strength and
constant dilaton.

Note that the results of subsection \ref{dimred} show that the
oxidation of the $D=5$ null solutions leads to $D=6$ solutions with
$u$-dependent Gibbons-Hawking base space. This suggests that it might
be possible to extend the analysis of the present subsection to
include $u$-dependence, although we shall not do so here.

\subsubsection*{Example: black string}

Introduce spherical polar coordinates
$(R,\theta,\phi)$ for ${\bf x}$ on the base space, and take
\be
 H_2 = \frac{1}{R}, \qquad \chi_i dx^i = \cos \theta d\phi.
\ee
The base space is then flat: let $R=\frac{1}{4} \rho^2$ to get
\be
 ds_4^2 = d\rho^2 + \frac{\rho^2}{4} \left(
 (\sigma_1)^2+(\sigma_2)^2+(\sigma_3)^2 \right),
\ee
where $\sigma_i$ are left-invariant 1-forms on $SU(2)$ - see \cite{d5}
for details. In the above notation, we have $\sigma=\sigma_3 = d\psi +
\cos \theta d\phi$ where $\psi=z$. $(\theta,\phi,\psi)$ are Euler
angles on $S^3$. We shall look for a solution for which all of
the harmonic functions are described by monopole sources at
the origin $R=0$. Furthermore, assume that the solution is
asymptotically flat as $R \rightarrow \infty$, i.e., $H \rightarrow
1$ and $\beta, \omega \rightarrow 0$. The only way of ensuring
$\beta_0 \rightarrow 0$ is to take $H_3 = 0$, which implies that
$\bbeta=0$ (so the solution is non-twisting). Now also assume that
$\bomega=0$. Then, integrating equation \reef{eqn:omega0} and demanding
$\omega_0 \rightarrow 0$ implies that $H_4 \propto H_2$. However, from
the definition of $H_4$ (equation \reef{eqn:csol}) we see that $H_4$ is
arbitrary up to addition of a multiple of $H_2$, so we can 
 set $H_4=0$. Doing so we arrive at the solution
\be
\label{eqn:stringsol}
 H = 1 + \frac{\mu}{\rho^2}, \qquad {\cal F} = - 1 - \frac{p}{\rho^2}, \qquad \beta=0,
 \qquad \omega = \frac{j}{2 \rho^2} \sigma_3,
\ee
where $\mu$, $p$ and $j$ are constants. The constant term in ${\cal
 F}$ can be adjusted by shifting $v \rightarrow v + c u$; the above
choice will be convenient below. This solution describes a
rotating momentum-carrying string. As we shall see below, it has a
regular event horizon at $\rho=0$. When oxidized to $D=11$, this
solution is a special case of a 4-charge solution constructed
in \cite{cvetic:97}.

\sect{Solutions with a horizon}

\label{sec:horizon}

\subsection{Gaussian null coordinates}

Supersymmetric solutions with event horizons are of special
interest. If a solution has an event horizon then it must be preserved
by $V$ hence $V$ must be tangent to the horizon, so the event horizon
is a Killing horizon of $V$. In this section we shall consider
all solutions with a Killing horizon of $V$.

We shall start by choosing a suitable gauge, corresponding to Gaussian
null coordinates adapted to a Killing horizon ${\cal H}$
\cite{friedrich:99,reall:02}. Pick a partial Cauchy surface $\Sigma$. Let ${\cal H}_0$
denote the intersection of $\Sigma$ with the horizon. Introduce
coordinates $x^m$ on ${\cal H}_0$. Define a coordinate $v$ on ${\cal H}$ to be
the parameter distance of a point from ${\cal H}_0$ along the orbits of
$V$. Let $n$ be the unique (past directed) null vector on ${\cal H}$ that obeys $V
\cdot n = -1$ and $V \cdot X=0$ for any vector $X$ tangent to surfaces
of constant $v$ in ${\cal H}$. Consider the null geodesic from a point
$p \in {\cal H}$ with tangent $n$. Let the coordinates of a point
affine parameter distance $r$ along this geodesic be $(v,r,x^m)$,
where $(v,x^m)$ are the coordinates of $p$. The line element must take
the form
\be
 \label{eqn:horizon}
 ds^2 = - 2 dr dv - 2 r h_m (r,x) dv dx^m - \gamma_{mn}(r,x) dx^m dx^n,
\ee
with $V = \partial/\partial v$, the horizon is at $r=0$ and $h_m$ and
$\gamma_{mn}$ must be smooth functions of $r$ in a neighbourhood of
the horizon $r=0$.

As an example, consider the string solution \reef{eqn:stringsol}:
\ba
 ds^2 &=& \left( 1 + \frac{\mu}{\rho^2} \right)^{-1} \left[2 du' dv' +
 \frac{j}{\rho^2} du' \sigma_3 - \left(1 + \frac{p}{\rho^2} \right)
      {du'}^2 \right] \nn
 &-& \left(1 +
 \frac{\mu}{\rho^2} \right) \left[ d\rho^2 + \frac{\rho^2}{4} \left(
   (\sigma_1)^2 + (\sigma_2)^2 + (\sigma_3)^2 \right) \right].
\ea
We have included primes on some coordinates to avoid confusion with
the coordinates of \reef{eqn:horizon}. It is convenient to identify $u'
\sim u' + l$ to render the length of the string finite. Consider
a coordinate transformation defined by
\be
 dv' = dv - A(r) dr, \qquad du' = du - B(r) dr, \qquad d\psi = d\psi' -
 C(r) dr, \qquad \rho d\rho = 2 \sqrt{D(r)} dr,
\ee
where $u \sim u+l$,
and choose the functions $A,B,C,D$ so that the metric takes the form
of \reef{eqn:horizon}. One obtains
\be
 D = \left(1+\frac{p}{\rho^2} \right)\frac{\rho^2}{4} - \left( 1 +
 \frac{\mu}{\rho^2} \right)^{-2} \frac{j^2}{4\rho^4}.
\ee
This is required to be positive, which implies
\be
\label{eqn:noctc}
 p > \frac{j^2}{\mu^2}.
\ee
For small $\rho$ we have
\be
 r = \frac{1}{2} \left( p - \frac{j^2}{\mu^2} \right)^{-1/2} \rho^2 + {\cal
   O}(\rho^4).
\ee
In the form \reef{eqn:horizon}, the solution has
\be
 r h_m dx^m = - \rho^2 (\rho^2 + \mu)^{-1} du,
\ee
\ba
 \gamma_{mn}(r,x) dx^m dx^n &=& \frac{(\rho^2+p)}{(\rho^2 + \mu)} \left( 1-
 \frac{j^2}{(\rho^2+p)(\rho^2 +\mu)^2}\right) du^2 \nn &+& \frac{1}{4}
 (\rho^2+\mu) \left[ \left(d\psi' + \cos \theta d\phi - \frac{2
     j}{(\rho^2+\mu)^2} du \right)^2 + d\theta^2 + \sin^2\theta d\phi^2
   \right] .
\ea
For fixed $r$ (i.e. fixed $\rho$), a shift $\psi' \rightarrow \psi' +
cu$ shows that this is just  the standard product metric on $S^1
\times S^3$. The sizes of the $S^1$ and $S^3$ vary with $r$,
approaching constant non-zero values as $r \rightarrow 0$ or $r
\rightarrow \infty$.

Since $\gamma_{mn}$ and $h_m$ are smooth at $r=0$, we have shown that
this solution must have a regular horizon there. The condition
\reef{eqn:noctc} ensures that the identification of $u$ is consistent
with a regular horizon. Upon dimensional reduction, this solution
gives a rotating black hole in $D=5$, with \reef{eqn:noctc} being the
condition for this black hole to have a regular horizon instead of
naked closed timelike curves
\cite{bmpv,tseytlin:96a,gauntlett:99}. For example, if $p=\mu$ then
the black hole is a solution of minimal $D=5$ supergravity and
\reef{eqn:noctc} reduces to $j^2 < \mu^3$, as obtained in
\cite{gauntlett:99}.

\subsection{Near-horizon geometry}

It has recently been realized that the black hole uniqueness theorems
do not extend to higher dimensions \cite{emparan:02}. However,
in \cite{reall:02}, a uniqueness theorem was proved for {\it supersymmetric}
black hole solutions of minimal $D=5$ supergravity. It is
of interest to ask whether a uniqueness theorem can also be proved
for supersymmetric black {\it strings} in $D=6$. We shall not attempt
that here, but will instead just repeat the first step of
\cite{reall:02}, namely to determine all possible near-horizon
geometries for supersymmetric solutions with a spatially compact Killing horizon,
i.e., compact ${\cal H}_0$. The black string solution of the previous
subsection is an example of such a solution but note that there exist black
strings with regular horizons for which ${\cal H}_0$ {\it cannot} be
rendered compact by making identifications. A simple example would be
the solution of the previous subsection with $p=j=0$. The naive
identification of this solution destroys regularity of the horizon
\cite{gibbons:94}.

The Gaussian null coordinates of \reef{eqn:horizon} give a coordinate
system of the form discussed in section \ref{subsec:coords} if we
identify $u=-r$, $\beta_m=u h_m$, $\omega=0$, ${\cal F}=0$
and $H=1$. Hence it is straightforward to apply our general formalism
to supersymmetric solutions with horizons. For example, one can
easily read off the field strength for such a solution from the general case.
To determine the possible near-horizon geometries, we just have to
evaluate everything on ${\cal H}_0$, i.e., at $r=0$.
Evaluating equation \reef{eqn:biana} at $r=0$ gives
\be
\label{eqn:coclosed}
\tilde{d} \star_4 h = 0 \qquad \mbox{on ${\cal H}_0$.}
\ee
Equation \reef{eqn:cdJ} becomes
\be
\label{eqn:hypercomp}
 \tilde{d} J^i = h \wedge J^i \qquad \mbox{on ${\cal H}_0$.}
\ee
This implies that the 2-forms $J^i$ form an integrable  hyper-hermitian
structure on ${\cal H}_0$ with Lee form $h$ (see e.g. \cite{gt}) i.e. each of the three
almost complex structures is integrable. The integrability condition of eqution
\reef{eqn:hypercomp}
tells us that $\tilde{d} h$ is self-dual on ${\cal H}_0$, which also
follows from equation \reef{eqn:betasd}.

The special case in which $h$ vanishes on ${\cal H}_0$ is straightforward. In
this case, ${\cal H}_0$ must be a compact hyper-K\"ahler space and
hence either $\bT^4$ or $K3$. The near-horizon limit of the solution is
just the product $\bR^{1,1} \times {\cal H}_0$ with vanishing flux.

Now we shall assume that $h \ne 0$ on ${\cal H}_0$, with ${\cal H}_0$
compact. In this case, integrating
$\tilde{d}h \wedge *_4 \tilde{d}h$ over ${\cal H}_0$ using self-duality
and Stokes' theorem yields the result
\be
 \tilde{d} h = 0 \qquad \mbox{on ${\cal H}_0$}.
\ee
Hence $h$ is closed and co-closed on ${\cal H}_0$. It follows that
\be
\label{eqn:integral}
 I \equiv \int_{{\cal H}_0} \tilde{\nabla}_m h_n \tilde{\nabla}^m h^n = - \int_{{\cal H}_0}
 R_{mn} h^m h^n,
\ee
where $\tilde{\nabla}$ is the metric connection associated with $\gamma_{mn}$,
indices have been raised with $\gamma^{mn}$, and $R_{mn}(x)$ is the
Ricci tensor of ${\cal H}_0$. This can be obtained from an
integrability condition for \reef{eqn:hypercomp} but it is more easily
obtained by the following trick. Note that (since $h$ is closed)
locally we can write $h = -2 \Omega^{-1}
\tilde{d} \Omega$, and one then sees that $\Omega^{2} J^i$ yield an
integrable hyper-K\"ahler structure with metric $\bar{\gamma}_{mn} =
\Omega^{2} \gamma_{mn}$. Since this metric must be Ricci flat, we can
therefore calculate the Ricci tensor for $\gamma_{mn}$ in terms of
$\Omega$. Now $\Omega$ is only defined locally, but $h$ is defined
globally so we must rewrite the expression for the Ricci tensor using
only $h$. The result is
\be
 R_{mn} = - \tilde{\nabla}_m h_n - \frac{1}{2} h_m h_n + \frac{1}{2} h_p h^p
 \gamma_{mn} - \frac{1}{2} \tilde{\nabla}_p h^p \gamma_{mn}~,
\ee
which is symmetric in $m$ and $n$ because $\tilde{d} h = 0$ on
${\cal H}_0$. In the present case, the final term vanishes because of
 \p{eqn:coclosed}. Plugging this into equation \reef{eqn:integral} and integrating by
parts yields $I=0$ hence
\be
 \tilde{\nabla}_m h_n = 0 \qquad \mbox{on ${\cal H}_0$}.
\ee
So ${\cal H}_0$ admits a covariantly constant vector $h$. Now define a constant
$L$ by $4L^{-2} = h_m h^m$ and define a coordinate $\alpha$ on ${\cal H}_0$ to
be the parameter along the integral curves of $h$, so $h =
 \partial/\partial \alpha$. The metric on ${\cal H}_0$ must be
\be
 ds_4^2 = 4 L^{-2} \left( d\alpha + \nu \right)^2 + \gamma_{ij} dx^i
 dx^j,
\ee
where $\nu$ and $\gamma_{ij}$ are independent of
 $\alpha$. $\tilde{d}h=0$ implies that locally we can write $\nu =
 \tilde{d} \lambda$, which can be gauged away by shifting
 $\alpha$. Examining $R_{mn}$ reveals that the Ricci tensor
 of $\gamma_{ij}$ is $R_{ij} = 2 L^{-2} \gamma_{ij}$ hence
 $\gamma_{ij}$ must be locally isometric to the metric on a round
 $S^3$ of radius $L$. So locally we have
\be
 ds_4^2 = L^2 \left( dZ^2 + d\Omega^2 \right),
\ee
where we have performed a change of coordinates $\alpha = - (L^2/2) Z$,
which gives $h = - 2 dZ$ on ${\cal H}_0$. So the metric on ${\cal H}_0$ is locally
isometric to the standard metric on $S^1 \times S^3$.

Plugging these results back into the six dimensional metric yields
\be
 ds^2 = - 2 dr dv + 4 r dv dZ - L^2 \left( dZ^2  +  d\Omega^2 \right) +
 {\cal O}(r^2) dv dx^m + {\cal O}(r) dx^m dx^n,
\ee
where $x^m = \{Z,\Omega\}$. Taking the near horizon limit $r =
\epsilon \hat{r}$, $v = \hat{v}/\epsilon$, $\epsilon \rightarrow
0$, and making the change of coordinates $\hat{r}=\hat{u}e^{2Z}$ gives
\be
 ds^2 = - 2e^{2Z} d\hat{u} d\hat{v} - L^2 \left( dZ^2  +  d\Omega^2 \right),
\ee
which is the metric of $AdS_3 \times S^3$. Of course, this result is
only local since our discussion of the geometry of ${\cal H}_0$ was
purely local. Globally, the solution will be some identification of
$AdS_3 \times S_3$, the simplest possibility being just $Z \sim Z+{\rm
constant}$. Note that, in these coordinates, the horizon we are
studying is the one at $\hat{u}=0$ (not $Z=-\infty$ since compactness of
${\cal H}_0$ implies that $Z$ is bounded).

In summary, we have shown that any supersymmetric solution of minimal
six dimensional supergravity with a spatially compact Killing horizon of $V$
must have a near-horizon geometry that is $\bR^{1,1} \times \bT^4$,
$\bR^{1,1} \times K3$, or identified $AdS_3 \times S^3$.

\sect{Maximal Supersymmetry}

\label{sec:max}

In order to determine the maximally supersymmetric solutions of the theory, we observe that
the integrability conditions \reef{eqn:integ} imply that
\be
\label{eqn:reicurv}
R_{\nu \mu \rho \lambda} - \nabla_\nu G_{\mu \rho \lambda}+
\nabla_\mu G_{\nu \rho \lambda} +2 G_{\mu \alpha [\rho |} G_\nu{}^\alpha{}_{|\lambda]} =0 \ .
\ee
Hence, on antisymmetrizing on the indices $\mu$, $\rho$, $\lambda$ and making use of
({\ref{eqn:pluck}}) together with $dG=0$, it is straightforward to show that
\be
\label{eqn:parg}
\nabla G=0
\ee
i.e. $G$ is parallel with respect to the Levi-Civita connection. Substituting this into
({\ref{eqn:reicurv}}) we obtain
\be
\label{eqn:reicurvb}
R_{\nu \mu \rho \lambda} =-2 G_{\mu \alpha [\rho |} G_\nu{}^\alpha{}_{|\lambda]} \ .
\ee
Observe, that as $G$ is parallel, so is the Riemann tensor and hence the geometry must be locally
symmetric. In addition, ({\ref{eqn:pluck}}) implies that $G$ satisfies an orthogonal
Pl\"ucker-type relation. Hence, as a consequence of section 2.2 in \cite{pluck}, it is straightforward
to show that we can write
\be
\label{eqn:gdec}
G= P+ \star P
\ee
where $P$ is a decomposable 3-form. In addition $\nabla G=0$ implies that $\nabla P=0$.
To proceed we shall make a modification of the reasoning used in section 3.3 in \cite{maxi}.
There are two cases to consider.

In the first case, the 3-form $P$ is not null. Then $P$
induces a local decomposition of the manifold into a product of two three dimensional
symmetric spaces $M=M_1 \times M_2$ with $P \propto d\mathrm{vol} (M_1)$ and $\star P \propto
d \mathrm{vol} (M_2)$. Without loss of generality we can assume that $P$ has positive norm and so
$M_1$ is Lorentzian (with mostly minus signature). Then we have
\be
G = \chi \big[d\mathrm{vol} (M_1) + d\mathrm{vol} (M_2) \big]
\ee
for constant $\chi$. If $\chi=0$ then it is clear that the Riemann tensor vanishes and
hence the geometry is flat. Otherwise, if $\chi \neq 0$, then
on $M_1$ the components of the Riemann curvature tensor satisfy
\be
R_{ijmn}= \chi^2 (g_{im} g_{jn}-g_{jm} g_{in})
\ee
so $M_1$ is isometric to $AdS_3$, and on $M_2$ the components of the
Riemann curvature tensor satisfy
\be
R_{ijmn}= -\chi^2 (g_{im} g_{jn}-g_{jm} g_{in})
\ee
so $M_2$ is isometric to $S^3$. Both the $AdS_3$ and $S^3$ have the same radius of curvature.

In the second case, the 3-form $P$ is null. It is known from \cite{cwpaper}
that all Lorentzian symmetric 6-manifolds admitting parallel null forms are
locally isometric to a product $M= CW_d (A) \times Q_{6-d}$ for $d=3,4,5,6$ where $Q_{6-d}$ is a
Riemannian symmetric space and $CW_d (A)$ is a $d$-dimensional Cahen-Wallach space.
As $P$ is null and decomposable, we must have
\be
P= dx^- \wedge \psi
\ee
where $dx^-$ is a parallel null form which exists in every Cahen-Wallach space and
$\psi$ is a parallel 2-form on $M$ with negative norm. It is straightforward to see that the
components of the Riemann curvature of $Q$ must all vanish and hence $Q_{6-d}= \bR^{6-d}$.
The metric on $M= CW_d (A) \times \bR^{6-d}$ can be written locally as
\be
ds^2 =2 dx^+ dx^- + \sum_{i,j=1}^4 A_{ij} x^i x^j (dx^-)^2-\sum_{i=1}^4 (dx^i)^2
\ee
where $A$ is a symmetric $4 \times 4$ matrix with constant coefficients, which is degenerate along the
$\bR^{6-d}$ directions. We can choose co-ordinates on $\bR^4$ so that
\be
P= \mu dx^- \wedge dx^1 \wedge dx^2 \ .
\ee
The maximally supersymmetric Cahen-Wallach type solutions have been examined  in
\cite{meesnote} and it is straightforward to show that the only possible solution
is in fact $CW_6 (A)$ with $A_{ij}=\mu^2 \delta_{ij}$.

To summarize, the only maximally supersymmetric solutions of the minimal
six-dimensional supergravity are $\bR^{1,5}$, $AdS_3 \times S^3$
and the $CW_6$ solution described above.

\sect{Outlook}

\label{sec:outlook}

We have presented a general form for all supersymmetric 
solutions of minimal supergravity in six dimensions. The solutions   
preserve either half or all of the supersymmetry. Our method relies 
on the analysis of the algebraic and differential constraints obeyed by 
certain differential forms constructred as spinor bilinears, and is related to
the mathematical notion of $G$-structures. Our results, together with those
of \cite{d5} and \cite{GG} which analysed minimal supergravities in five 
dimensions, provide encorouging evidence that our approach could be extended
to other supergravity theories. 

For instance, we expect that minimal  $N=2$, $D=4$ {\em gauged} supergravity 
could be easily analysed using our methods. Recall that the ungauged theory 
was tackled some time ago by Tod \cite{tod:83}, from which it is known 
that there are a timelike and a null case to analyse. In the gauged theory 
the algebraic structure will remain unchanged while new differential
conditions will arise. 

Similarly, minimal $D=6$ gauged supergravity could be analysed by 
generalizing our results for the ungauged theory. Here one adds
the tensor multiplet mentioned at the beginning of section
\ref{sec:basics}, and a vector multiplet whose 
bosonic field is a one-form potential $A$. From a Killing spinor one
can construct a vector $V$ and 3-forms $X^i$ obeying the same
algebraic relations as in the present paper, although the differential
relations will be different. Given the results
of \cite{GG} one anticipates that the resulting $SU(2)\ltimes \bR^4$ structure
will be some generalization of the one encountered here. A systematic analysis 
of this theory might address some of the questions recently raised in
\cite{guven:03}.  

More generally, it is interesting to ask which combinations of vector
and tensor multiplets can be added to the minimal theories for them to
remain tractable using our techniques. Dimensional
reduction of the minimal $D=6$ theory yields the minimal $D=5$ theory
coupled to a vector multiplet, hence all supersymmetric solutions of
the latter theory must arise as a subset of the solutions presented
here. So the case of a single vector multiplet is certainly
tractable. Similarly, reduction to $D=4$ yields the minimal $N=2$
$D=4$ theory coupled to 3 vector multiplets, so this theory should
also be tractable. These examples suggest that it might be fruitful to
examine the cases in which arbitrary many vector multiplets are
present. More ambitiously, one might hope that a similar analysis
could be applied to non-abelian gauged supergravities, which in recent years 
have proved valuable tools for finding new solutions of interest in string
theory.

The results of \cite{GP} and \cite{GMW} have shown that the same techniques
prove useful in classifying and analyzing supersymmetric solutions of 
higher dimensional supergravities. In 
particular in \cite{GP} the most general form of supersymmetric solutions
admitting at least a ``timelike'' Killing spinor in $D=11$ supergravity was
given, while in \cite{GMW} static solutions of $D=10$ Type II theories with NS
fields were analysized in detail.  Although in these theories the form of the
solutions is determined somewhat implicitly, it is nevertheless useful 
to have the most general solutions catalogued. To complete such a
catalogue, one would have to examine null solutions, which in general
preserve 1/32 supersymmetry. The null Killing vector field will, as in $D=6$,
generally be twisting, so we hope that the analysis of such solutions
presented here will be of some use in understanding how things work in
$D=10,11$.

\vspace{4.0in}

\begin{center} {\bf Acknowledgments} \end{center}

We thank Jerome Gauntlett, Chris Hull, Arkady Tseytlin and Dan Waldram for
useful discussions. J.~G.~ was supported by EPSRC.
D.~M.~ was supported by an EC Marie Curie Individual Fellowship under contract
number HPMF-CT-2002-01539. H.~S.~R.~ was supported by PPARC.

\appendix
\makeatletter
\renewcommand{\theequation}{A.\arabic{equation}}
\@addtoreset{equation}{section} \makeatother

\section{Conventions and useful identities}
\label{conventions}

We follow the spinor conventions of \cite{NS1,NS2}. The $8 \times 8$ Dirac matrices
in six dimensions obey the Clifford algebra
\bea
\{\gamma_\alpha,\gamma_\beta \} = 2 \eta_{\alpha \beta}~, \qquad
\eta_{\alpha \beta} = (+,-,-,-,-,-)~.
\eea
where $\alpha, \beta,\ldots$ are tangent space indices. Curved indices
will be denoted by $\mu,\nu,\ldots$.
The conjugation matrix $C$ is symmetric and can be set to unity. Hence, in
this representation the $\gamma$ matrices are antisymmetric
\bea
\gamma_\alpha^T & = & -\gamma_\alpha~.
\eea
The chirality projector is defined as
\bea
\gamma_7  =  \gamma_0 \gamma_1 \cdots \gamma_5~, \qquad
\gamma_7^2=1~,\quad\gamma_7^T=-\gamma_7~.
\eea
The duality relation of the $\gamma$ matrices reads
\bea
\label{dualityofgammas}
\gamma^{\alpha_1\dots \alpha_n} =\frac{(-1)^{[n/2]}}{(6-n)!}\e^{\alpha_1\dots \alpha_n
\beta_1\dots \beta_{6-n}}\gamma_{\beta_1\dots \beta_{6-n}}\gamma_7~,
\eea
with $\epsilon^{012345}=+1$. All the spinors are symplectic Majorana
\bea
\chi^A = \e^{AB}\bar{\chi}^T_B~, \qquad \bar{\chi}_A=(\chi^A)^\dagger
\gamma_0
\eea
which means that $\bar{\chi}^A = \chi^{AT}$~.

For any given four symplectic Majorana-Weyl spinors $\psi_1,\dots, \psi_4$ with
chiralities $\gamma_7\psi_2=c_2\psi_2$, $\gamma_7\psi_4=c_4\psi_4$,
the Fierz rearrangement formula reads
\bea
\bar{\psi}_1\psi_2  \bar{\psi}_3\psi_4 & = &
\frac{1}{8}(1+c_2c_4)\left[ \bar{\psi}_1\psi_4
\bar{\psi}_3\psi_2    -\frac{1}{2}\bar{\psi}_1\gamma^{rs}\psi_4
\bar{\psi}_3\gamma_{rs}\psi_2  \right]\nn
&&\frac{1}{8}(1-c_2c_4)\left[ \bar{\psi}_1\gamma^r\psi_4
\bar{\psi}_3\gamma_r\psi_2    -\frac{1}{12}\bar{\psi}_1\gamma^{rst}\psi_4
\bar{\psi}_3\gamma_{rst}\psi_2  \right]~.
\label{Fierz}
\eea
Note that there is a change in
sign with respect to \cite{NS2} because we are using
{\em commuting} spinors.
Since $C=1$ we can use $\gamma_0$ as the intertwing operator between a
representation of gamma matrices and its complex conjugate, in
particular
\bea
\gamma_0 \gamma_\alpha \gamma_0 & = & - \gamma_\alpha^*~.
\eea
Notice that this is indeed consistent with the simplectic Majorana condition,
which in $Sp(1)$ components can be written as
\be
\chi^1 = - \gamma_0 \chi^{2*}, \qquad \chi^2  =  \gamma_0 \chi^{1*}~.
\ee
From these we obtain the following reality properties of spinor bi-linears
\bea
(\bar{\e}^1\gamma_{\alpha_1\dots \alpha_n}\e^2)^* & = & (-1)^n
\bar{\e}^2\gamma_{\alpha_1\dots \alpha_n}\e^1 \nn
(\bar{\e}^1\gamma_{\alpha_1\dots \alpha_n}\e^1)^* & = & (-1)^{n+1}
\bar{\e}^2\gamma_{\alpha_1\dots \alpha_n}\e^2~.\label{reality}
\eea
We also note the following useful gamma-matrix identity
\bea
\label{eqn:gamident}
\gamma_\alpha \gamma^{\beta \gamma \delta} \gamma^\alpha & = & 0~.
\eea

\subsubsection*{Algebraic identities satisfied by spinor bi-linears}

The bi-linears $V$ and $\Omega^{AB}$ defined in the main text in
(\ref{defV}) and (\ref{defO}) satisfy some lengthy but useful
relations which arise from the Fierz identity
(\ref{Fierz}). Setting $\psi_1 = \epsilon^A$, $\psi_2 =
\gamma_\lambda \epsilon^B$, $\psi_3 = \epsilon^C$, $\psi_4 =
\gamma_\rho \epsilon^D$ in (\ref{Fierz}) we obtain
\bea
\label{eqn:fierza}
(\epsilon^{AB} \epsilon^{CD}-{1 \over 2} \epsilon^{AD} \epsilon^{CB})
V_\lambda V_\rho +{1 \over 4} \epsilon^{AD} \epsilon^{CB} V_\mu V^\mu
g_{\lambda \rho} = -{1 \over 8} \Omega^{AD \ \mu \nu}{}_\rho
\Omega^{CB}{}_{\mu \nu \lambda}
\nn
-{1 \over 4} \epsilon^{CB}
V^\mu \Omega^{AD}{}_{\mu \lambda \rho}+{1 \over 4} \epsilon^{AD}
V^\mu \Omega^{CB}{}_{\mu \lambda \rho}~.
\eea
Contracting this with $g^{\lambda \rho}$ and setting
$A=B$, $C=D$ we find $V_{\mu} V^{\mu}=0$, i.e., $V$ is null.
Now use ({\ref{eqn:fierza}}), setting $A=B$, $C=D$, to  obtain
\bea
{1 \over 2} (\epsilon^{AB})^2 V_\lambda V_\rho = -{1 \over 8}
\Omega^{AB \ \mu \nu}{}_\rho \Omega^{AB}{}_{\mu \nu \lambda}
+{1 \over 2} \epsilon^{AB} V^\mu \Omega^{AB}{}_{\mu \lambda \rho}~.
\label{eqVV}
\eea
By anti-symmetrizing this on $\lambda \ , \rho$ we obtain
\be
i_V \Omega^{AB}=0.
\ee
Next use $\psi_1 = \gamma_\nu \gamma^{\rho \sigma}
\epsilon^A$,
$\psi_2 = \epsilon^B$, $\psi^3 = \epsilon^C$, $\psi_4 = \gamma^\nu
\gamma^{\lambda \mu} \epsilon^D$ in the Fierz identity;
note that $\psi_2$ and $\psi_4$ are of opposite chirality. Using
(\ref{eqn:gamident}) we obtain
\bea
\label{eqn:compalg}
\Omega^{AB \ \rho \sigma}{}_\nu \Omega^{CD \ \nu \lambda \mu} &=&
(2 \epsilon^{CB} \epsilon^{AD}- \epsilon^{AB} \epsilon^{CD})
\big( g^{\sigma \lambda}V^\mu V^\rho + g^{\mu \rho}V^\lambda V^\sigma
-g^{\sigma \mu}V^\rho V^\lambda
-g^{\rho \lambda} V^\mu V^\sigma \big)
\nn
&-& \epsilon^{AB} \big(\Omega^{CD \ \sigma \lambda \mu}V^\rho - \Omega^{CD \
\rho \lambda \mu}V^\sigma \big)
-\epsilon^{CD} \big( \Omega^{AB \ \rho \sigma \lambda}V^\mu - \Omega^{AB \ \rho
\sigma \mu}V^\lambda \big)
\nn
&-&2 \epsilon^{CB} \big(\Omega^{AD \ \rho \lambda \mu}V^\sigma - \Omega^{AD \
\sigma \lambda \mu}V^\rho \big)~.
\eea

\sect{Integrability conditions}

\makeatletter
\renewcommand{\theequation}{B.\arabic{equation}}
\@addtoreset{equation}{section} \makeatother

Note that the Killing spinor equation
\be
\nabla_\mu \epsilon = {1 \over 4} G_{\mu \rho \lambda} \gamma^{\rho \lambda} \epsilon
\ee
for self-dual $G$ implies the following integrability condition
\be
\label{eqn:integ}
\big( R_{\nu \mu \rho \lambda} - \nabla_\nu G_{\mu \rho \lambda}+
\nabla_\mu G_{\nu \rho \lambda} +2 G_{\mu \sigma \rho} G_\nu{}^\sigma{}_\lambda \big)
\gamma^{\rho \lambda} \epsilon =0
\ee
On contracting this identity with $\gamma^\mu$, and using the self-duality of $G$, we
obtain
\be
{1 \over 3} dG_{\mu \nu \rho \lambda} \gamma^{\mu \rho \lambda} \epsilon
+ \big(-2 R_{\nu \lambda} -2 \nabla^\mu G_{\mu \nu \lambda} +2 G_{\nu \mu \rho}
G_\lambda{}^{\mu \rho} \big) \gamma^\lambda \epsilon =0
\ee
where we note that as a consequence of the self-duality of $G$,
\be
\label{eqn:pluck}
G_{\mu [\nu \rho} G^\mu{}_{\lambda] \sigma}=0 ~.
\ee
On imposing the Bianchi identity $dG=0$ we obtain
\be
\label{eqn:einint}
E_{\mu \nu} \gamma^{\mu} \epsilon =0
\ee
where
\be
E_{\mu \nu} \equiv R_{\mu \nu} - G_{\mu \rho \sigma} G_\nu{}^{\rho \sigma}
\ee
In particular, we observe that ({\ref{eqn:einint}}) implies that, in the null basis,
$E_{- \alpha}=0$. In addition, ({\ref{eqn:einint}}) also implies that
\be
E_{\mu}{}^\nu E_{\mu \nu} =0
\ee
with no sum on $\mu$, from which we also find that $E_{+a}=E_{ab}=0$. Hence, the integrability of the
Killing spinor equation is sufficient to imply that all except the $++$ components of the Einstein
equation hold automatically.

\sect{Spin connection}

\label{sec:spinconn}

The spin connection is defined by
\be
 \omega_{\mu \alpha \beta} = e^{\nu}_{\alpha} \nabla_{\mu} e_{\beta
   \nu}.
\ee
In the basis \p{eqn:basis}, the components of the spin connection are
given by \p{eqn:minuscompts} and
\be
 \omega_{++a} = H \left( \frac{1}{2} \cD \cF - \frac{1}{2} \cF
 \dot{\beta} - \dot{\omega} \right)_a,
\ee
\be
 \omega_{b+a} =  \frac{1}{2} \left( \cD \omega + \frac{\cF}{2} \cD \beta
   \right)_{ab} + \frac{1}{2} \partial_u (H h_{mn} ) \tilde{e}^m_a \tilde{e}^n_b,
\ee
\be
 \omega_{+ab} = - \frac{1}{2} \left( \cD \omega + \frac{\cF}{2}
 \cD \beta \right)_{ab} - H \tilde{e}^m_{[a} \partial_u \tilde{e}_{b]m},
\ee
\be
 \omega_{cab} =  - H^{-1/2} \tilde{\omega}_{cab} + H^{-1} (\cD H)_{[a}
     \delta_{b]c} + \frac{1}{2} H^{1/2} \left( (\beta \wedge \dot{\tilde{e}}_a
   )_{bc} - (\beta \wedge \dot{\tilde{e}}_c
   )_{ab}+ (\beta \wedge \dot{\tilde{e}}_b
   )_{ca} \right)
\ee
where $\tilde{\omega}_{cab}$ are the basis components of the spin connection of the base
manifold with indices lowered by $\delta_{ab}$, and similarly $\tilde{e}_a = \tilde{e}^a$.

\end{document}